\documentclass{aa}  
\usepackage{lscape}
\usepackage{graphicx}
\usepackage{txfonts}
\usepackage[figuresright]{rotating}
\begin{document}
\title{Reaching the boundary between \\ stellar kinematic groups and very wide
binaries}   
\titlerunning{The Washington Double Stars with the widest angular separations}
\subtitle{The Washington Double Stars with the widest angular separations}
\author{Jos\'e A. Caballero}
%
%
\institute{
Departamento de Astrof\'{\i}sica y Ciencias de la Atm\'osfera, Facultad de
F\'{\i}sica, Universidad Complutense de Madrid, E-28040 Madrid, Spain.
\email{caballero@astrax.fis.ucm.es}}
\date{Received 29 May 2009 / Accepted 12 August 2009}

\abstract
{} 
{I~look for and characterise very wide binaries and multiple systems with
projected physical separations larger than $s$ = 0.1\,pc, which is generally
believed to be a sharp upper limit to the distribution of wide binay
semimajor axes.}   
{I~investigated in detail 30 Washington Double Stars with angular separations
$\rho >$ 1000\,arcsec.
I~discarded 23 of them as probably unbound systems based on
discordant astrometry, photometry, spectral types, and radial velocities.
The remaining seven systems were subject to a comprehensive data
compilation and derivation (multi-wavelength photometry, heliocentric distance,
multiplicity, age, mass, metallicity, membership in young kinematic group).}  
{Of the seven very wide systems, six have projected physical separations larger
than the hypothetical cutoff at $s$ = 0.1\,pc; four of them have separations $s
>$ 0.2\,pc.
Although there are two systems in young kinematic groups (namely HD~136654 and
BD+32~2572 in the Hyades Supercluster, and AU~Mic and AT~Mic~AB in the
$\beta$~Pictoris moving group), there is not a clear prevalence of young systems
($\tau <$ 1\,Ga) among these very wide binaries.
Finally, I~compare the binding energies of the seven systems with those of other
weakly bound systems in the field.} 
{}
\keywords{astronomical data bases: miscellaneous -- stars: binaries: general --
stars: binaries: visual -- stars: kinematics}    
\maketitle
%

\section{Introduction}
\label{introduction}

Binaries can be classified by physical separation into {\em close} and {\em
wide} binaries.
Close binaries include e.g. spectroscopic, astrometric, interferometric,
eclipsing, cataclysmic, and semi-detached binaries, which have quite
differentiated astrophysical properties.
However, there is subjectiveness in the determination of the boundary between
close and wide binaries: 
while for some authors a wide binary is a detached pair with a physical
separation of a few tens solar radii, enough to avoid Roche lobe filling and
mass transfer after the main sequence, for others the stars in a wide binary
must be separated by at least 10$^3$--10$^4$\,AU
(0.005--0.05\,pc)\footnote{1\,pc $\approx$ 206\,264.806\,AU.}, depending on 
the total mass of the system.  

It remains an open question whether the definition of wide binaries must be
``stretched a little'' to include common proper motion pairs (pairs of stars
traveling together through space without any discernible relative orbital
motion; Batten 1973). 
For example, just in the very closest solar neighbourhood, the 
gravitational binding between {Proxima Centauri} (M5.5Ve) and {$\alpha$~Cen~A}
and~B (G2V+K2IV), a celebrated common proper motion ``pair'' (Innes 1915) with a
physical separation $r$ = 12\,000$\pm$600\,AU (0.058$\pm$0.003\,pc) and a long
expected orbital period ($P \gtrsim$ 0.9\,Ma), has been repeatedly questioned
(Vo\^ute 1917; Wertheimer \& Laughlin 2006 and references therein). 

Some authors have proposed that $\alpha$~Cen~A and~B are the brightest members
in a stellar kinematic group including Proxima Centauri (e.g. Anosova \& Orlov
1991). 
Stars in a stellar kinematic group share common origin and Galactic spatial
velocities ($UVW$) and are typically young, with Hyades-like or younger ages
($\tau \lesssim$ 600\,Ma -- Soderblom \& Mayor 1993; Montes et~al. 2001;
Zuckerman \& Song 2004).  
Youth may partly explain the existence of some {\em very wide} binaries (or {\em
very wide} common proper motion pairs), with physical separations of more than 
0.1\,pc.
The younger a wide (or very wide) binary in the Galactic disc is, the less time
it has had to encounter individual stars and giant molecular clouds, whose
gravity will eventually tear them apart (e.g. Bahcall \& Soneira 1981; Retterer
\& King 1982; Weinberg et~al. 1987; Saarinen \& Gilmore 1989; Poveda \& Allen
2004).   
For instance, the common proper motion ``pair'' {AU~Mic} and
{AT~Mic~A} and~B has one of the widest projected physical separations yet
measured, $s \sim$ 0.23\,pc, and belongs to one of the youngest stellar
kinematic groups, the $\beta$~Pictoris moving group ($\tau \sim$ 12\,Ma --
Zuckerman et~al. 2001; Ortega et~al.~2004).  

Another way for a very wide binary to avoid encounters in the Galactic disc
is to belong to the Galactic halo stellar population. 
Because of the large inclination of their orbits, halo stars spend most of their
lives far from the Galactic plane, where the probability of encountering
stars and molecular clouds is minimum. 
As another example, the system {HD~149414}~AB and {BD--03~3968B}
was the widest metal-poor ``binary'' in the imaging search by Zapatero Osorio \&
Mart\'{\i}n (2004).
Projected physical separation between both components is $s \sim$ 0.27\,pc.
Low metallicities, as that measured in the primary of the system (the F8
subdwarf HD~149414~AB has [Fe/H] $\sim$ --1.4), are typical of halo
population~II stars. 
The three systems ($\alpha$~Cen, AU~Mic, and HD~149414~AB) will be discussed
next. 

There is a sharp cutoff in the number of very wide binaries with physical
separations larger than 0.1\,pc, possibly dictated by dynamical evolution, as
stated in classical (Tolbert 1964; Kraicheva et~al. 1985; Abt 1988; Weinberg \&
Wasserman 1988; Weis 1988; Close et~al. 1990; Latham et~al. 1991; Wasserman \&
Weinberg 1991 and references above) and modern works (Allen et~al. 2000; Palasi
2000; Chanam\'e \& Gould 2004; L\'epine \& Bongiorno 2007; Makarov et~al. 2008).
My aim is to characterise and look for very wide binaries and multiple systems
with projected physical separations larger than $s$ = 0.1\,pc 
(2\,10$^4$\,AU). 
Some of these systems will be among the least bound ones and might help to
trace the boundaries between very wide binaries, common proper motion
pairs, and stellar kinematic groups on the point of being disrupted. 

In this this paper, I~start a programme of identifying and investigating
the widest common proper motion pairs with a detailed analysis of the binary and
multiple system candidates with angular separations larger than 1000\,arcsec
in the Washington Double Stars (WDS) catalogue (Mason et~al. 2001). 
It is expected that these large angular separations would translate into large
physical separations, of the order of a tenth of a parsec.
However. the membership in a proper motion pair of many doubles has not been
confirmed since their discovery dates (as soon as the 19th century in some
cases) and most of them were last characterised in the pre-{\em Hipparcos} era
(i.e. no accurate projected physical separations could be measured).

\section{Analysis and results}
\label{analysisandresults}

\subsection{Data retrieval}

\begin{figure}
\centering
\includegraphics[width=0.49\textwidth]{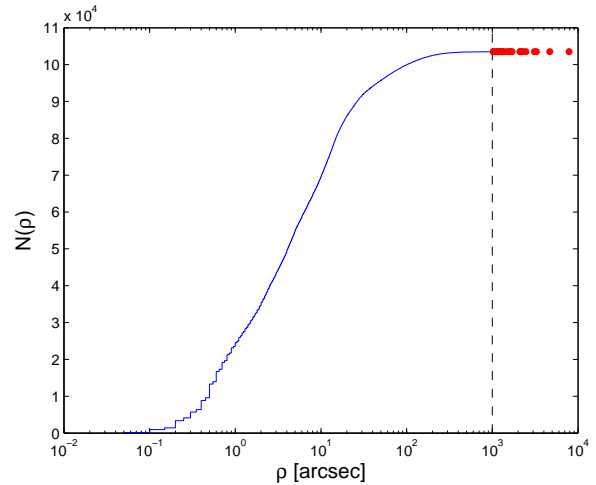}
\caption{Cumulative number of WDS pairs as a function of the angular
separation, $\rho$. 
Data points with $\rho >$ 1000\,arcsec (to the right of the vertical dashed
line) are for the 30 WDS binary candidates investigated here.}   
\label{xNvsrho}
\end{figure}

First, I compiled the angular separations, $\rho$, position angles, $\theta$,
coordinates, visual magnitudes, and identifiers of {104\,312} WDS pairs (as
in 2009 May). 
As can be seen in Fig.~\ref{xNvsrho}, the cumulative number of pairs
approximately increases with a power law in the interval $\rho \sim$ 0.2 to
20\,arcsec.
The distribution of angular separations is the \"Opik law (which formally
applies to the distribution of physical separations -- \"Opik 1924; Poveda \&
Allen 2004) folded with the distribution of observed systems with heliocentric
distance.  
The number of pairs increases at a more moderate rate from $\rho \sim$ 20
to 200\,arcsec and practically gets constant at larger separations.
Only 1\,\% [0.1\,\%] of the WDS pairs have angular separations $\rho >$
200\,arcsec [500\,arcsec].
There are {36} WDS pairs with tabulated separations $\rho >$ 1000\,arcsec
(actually, the WDS catalogue lists them as having $\rho \equiv$ 999\,arcsec). 
I~have carefully investigated the {36} of them, mainly using the Aladin sky
atlas (Bonnarel et~al.~2000).

Of the {36} WDS binaries, I~was not able to identify {five} binary
candidates:
\begin{itemize}

\item WDS~05463+5627 (LDS~3673; W.~J.~Luyten, proper motion catalogues).

\item WDS~04022+2808 (STF~481AD; F.~G.~W. Struve).
The primary in the system in the G8II triple star system 
\object{HD~25296}~ABC.

\item WDS~09510+0105 (GRV~1149; J.~Greaves, private communication).
The system might consist of the faint white dwarfs \object{WD~0948+013} and
\object{2QZ~J095234.0+011046}, but I~could not confirm it.

\item WDS~18382+2543 (BUP~185AC; S.~W.~Burnham, proper motion stars).
Coordinates have large uncertainties.

\item WDS~21435+2721 (A~299DE; R.~G.~Aitken).
Bright star \object{BD+26~4249} is likely the primary in system.

\end{itemize}

These unidenfications left only {31} WDS pairs to be investigated.
In Table~\ref{verywidebinaries.1}\footnote{NOTE to the readers:
in the astro-ph version, I~split the Table into two ones
(Tables~\ref{verywidebinaries.1} and~\ref{verywidebinaries.2}).}, I~list their
WDS identifier, discovery designation, Simbad name, equatorial coordinates,
proper motions, heliocentric distance, $V$, $J$, and $K_{\rm s}$ magnitudes,
spectral type, angular separation, and position angle.
Coordinates and $J$ and $K_{\rm s}$ magnitudes were homogeneously retrieved from
the Two Micron All Sky Survey (2MASS; Skrutskie et~al. 2006) except for the two
faint white dwarfs in the system WDS~02255-0904 (GRV~1148), whose coordinates
were taken from the sixth data release of the Sloan Digital Sky Survey (SDSS
DR6; Adelman-McCarthy et~al. 2008).
Proper motions were generally taken from the Tycho-2 catalogue (H{\o}g et~al.
2000), Salim \& Gould (2003), L\'epine \& Shara (2005 -- without tabulated error
bars), and R\"oser et~al. (2008), but in some particular cases (e.g. faint white 
dwarfs), the proper motions were borrowed from other sources in the literature
(e.g. Farihi et~al. 2005). 
Some of these values may be affected by observational errors.
I~retrieved parallaxes from the {\em Hipparcos} catalogue (van~Leeuwen 2007),
except for the white dwarf G~1--45~AB in the system WDS~01024+0504, for which
I~did it from van~Altena et~al. (1995)\footnote{The value of $d_\pi$ of
G~1--45~AB tabulated in the fourth edition of the general catalogue of
trigonometric parallaxes of van~Altena et~al. (1995) has a lower uncertainty
than the one in Harrington \& Dahn (1980), which was 22.2$\pm$2.0\,pc.}.
In hierarchical systems, I~used the parallax of the brightest component (e.g. 
$\alpha$~Cen~A for the primary in the system WDS~14396--6050).
$V$-band magnitudes of the brightest stars were taken from the original {\em
Hipparcos} catalogue (Perryman et~al. 1997).
For the faintest stars, if not available in other works (e.g. Weis 1984;
Ryan 1992), I~estimated the $V$-band magnitudes from four epochs of photographic
$B_J$ and $R_F$ magnitudes in the USNO-B1 catalogue (Monet et~al. 2003) or from
SDSS $g$ and $r$ magnitudes.
These estimations agreed with others (e.g. L\'epine \& Shara 2005).
Finally, I~tabulate spectral types as given by Simbad except for a few cases,
for which I~found more accurate determinations in the literature (e.g. Joy \&
Abt 1974).
Besides, I~compiled radial velocities from a number of works (Evans 1967;
Strassmeier et~al. 2000; Latham et~al. 2002; Nordstr\"om et~al. 2004; Torres
et~al. 2006).
Given the low fraction of stars with this measurement, I~did not list them in
Table~\ref{verywidebinaries.1}.

The stars in the system WDS~03330+0306 ({G~80--8} and
{NLTT~11184}) are actually separated by {\em less} than 1000\,arcsec
($\rho \approx$ 429.5\,arcsec) and, therefore, I~did not keep it in the next
analysis.

\subsection{Rejected binary candidates}

   \begin{table}
      \caption[]{Rejected wide binary candidates$^{a}$.} 
         \label{table.rejected}
     $$ 
         \begin{tabular}{l lllll l}
            \hline
            \hline
            \noalign{\smallskip}
WDS  		& $\mu$	&$d_{\rm phot}$	&$d_{\rm spec}$	& $d_\pi$	& V$_{\rm r}$	\\
identifier	& 	& 		& 		&		& 		\\
            \noalign{\smallskip}
            \hline
            \noalign{\smallskip}
00152+2454	& No:	& No		& ...		& ...	        & ...		\\
00400--1533	& No:	& No		& ...		& ...	        & ...		\\
00435+3351	& No	& No:		& No		& ...	        & ...		\\
00520+2035	& No	& No		& ...		& ...	        & ...		\\
01163--3217	& No:	& No		& ...		& ...	        & ...		\\
03442--6448	& No	& No		& No		& No	        & No		\\
07590--6338	& No	& No:		& ...		& ...	        & ...		\\
11125+3549	& No	& No		& No		& ...	        & ...		\\
13410+6808	& No:	& No		& No		& ...	        & ...		\\
13599+2520	& No	& No		& No		& ...	        & ...		\\
20084+1503	& No:	& No		& ...		& ...	        & ...		\\
20302+2651	& No	& No		& No		& ...	        & ...		\\
          \noalign{\smallskip}	
            \hline
         \end{tabular}
     $$ 
\begin{list}{}{}
\item[$^{a}$] Coordinates are provided in Table~\ref{verywidebinaries.1},
The colon after ``No'' indicates that the corresponding parameters of stars
in binary system candidates might be comparable if accounting for generous
error~bars. 
\end{list}
   \end{table}

I~classified the {30} WDS identified binaries with true angular
separations $\rho >$ 1000\,arcsec into confirmed and unconfirmed physical
systems based on their basic properties in Table~\ref{verywidebinaries.1}. 
First, I~made a preliminary filtering by rejecting {12} binary candidates
whose membership in a physically bound system is impractical because of their
different proper motions ($\mu$), heliocentric distances (photometric $d_{\rm
phot}$, spectroscopic $d_{\rm spec}$, or parallactic $d_\pi$), and/or radial
velocities (V$_r$), as summarised in Table~\ref{table.rejected}.  

\paragraph{WDS 03442--6448.}
The binary candidate formed by $\beta$~Ret~AB, a K2III spectroscopic binary, and
HD~24293, a G3V star, was proposed by Luyten in the mid-20th century based on
their similar proper motions.
They are bright stars and have accurate {\em Hipparcos} distance determinations.
The distances and true proper motions do not match, as well as their radial
velocities, as noted by Gliese \& Jahreiss (1988; precise radial velocities
are V$_{\rm r,K2III}$ = +50.8$\pm$0.9\,km\,s$^{-1}$, V$_{\rm r,G3V}$ =
+20.8$\pm$0.3\,km\,s$^{-1}$). 

\paragraph{WDS~00435+3351, 
WDS~00520+2035,
WDS~07590--6338,
WDS~11125+3549,
WDS~13599+2520,
WDS~20302+2651, and
WDS~23228+2208.}
The components in the {seven} binary candidates have different proper 
motions.  
These differences range between $\Delta \mu \approx$ 
44\,mas\,a$^{-1}$ for Giclas' WDS~00520+2035 
and over 400\,mas\,a$^{-1}$ for Weisse's WDS~00435+3351,
and cannot be explained by the different location in the sky or by multiplicity 
of one of the components (see the case of $\alpha$~Cen~AB and Proxima in 
Table~\ref{verywidebinaries.1}).
There is spectral type determination for the two components in four 
systems, and the secondaries have earlier spectral types than the primaries 
(i.e. the secondaries are located further away).

\paragraph{WDS 00152+2454,
WDS 00400--1533,
WDS 01163--3217, and
WDS 13410+6808.}
The secondaries of the {four} binary candidates are fainter and bluer than
their corresponding primaries.
For example, secondary G~57--17 in Giclas' WDS~11452+1821 is $\Delta J \approx$
3.70\,mag fainter, but $\Delta (V-J) \approx$ 1.90\,mag bluer, than primary
G~57--15.
From magnitudes, the secondaries seem to be normal dwarfs or subdwarfs with
earlier spectral types than the primaries.
In the case of Luyten's WDS~13410+6808, there is spectral type determination for
the two stars (Lee 1984):
the primary G~238--50 is an M2 dwarf at $d$ = 40$\pm$3\,pc, while the secondary
LP~40--200 is a K3 dwarf at $d \sim$ 100\,pc. 
Besides, proper motions of primaries and secondaries in the four systems
are different at the 3--5$\sigma$ level.

\subsection{Dedicated astro-photometric follow-up}
\label{dedicatedfollowup}

Next, I~performed an astro-photometric follow-up and investigated the possible
physical bounding of the remaining {18} binary candidates. 
Of them, {five} have hypothetical primaries and secondaries with reliable
published common proper motions and parallaxes from the {\em Hipparcos}
catalogue.
The hypothetical secondary in a {sixth} system, WDS~13090+3353, on the
contrary to the primary, does not have a {\em Hipparcos} measurement, but its
proper motion is tabulated in the accurate Tycho-2 catalogue.
In all six cases, the similarity between proper motions (and parallaxes)
of primaries and secondaries indicate that they are probably bound wide systems
(Section~\ref{probableboundsystems}).

The other {12} binary candidates were subject of a detailed proper motion
study. 
For each of them, I~collected precise coordinates of secondaries (and primaries,
if not too bright) at different astrometric epochs from SuperCOSMOS (Hambly
et~al. 2001) digitisations of the Palomar Observatory Sky Survey (POSS-I Red,
POSS-II Red, POSS-II Blue, POSS-II Infrared) and the 2MASS, and CMC14
(Carlsberg Meridian Catalogue; Evans et~al. 2002) catalogues.  
In a couple of cases, I was also able to use SDSS and Guide Star Catalog data.
With at least six astrometric epochs covering more than 45\,a, I~could measure 
new proper motions of {16} stars ({12} secondaries, {three} 
primaries, {one} tertiary) with unprecendented accuracy ($\delta \mu / \mu
\lesssim$ 1\,\%). 

I~display the proper motions of the components in the {12} followed-up 
systems in Table~\ref{newpropermotions}. 
Primaries and secondaries in {nine} of them have very different measured 
proper motions ($\Delta \mu \sim$ 25--90\,mas\,a$^{-1}$), and likely do not form
physically bound systems.
There is accurate SDSS photometry for at least one of them (WDS~11452+1821) that
supports this assumption (the $K_{\rm s}$ magnitudes and $g-K_{\rm s}$ colours
of the hypothetical primary and secondary are 8.260$\pm$0.016 and
6.071$\pm$0.016, and 12.097$\pm$0.023\,mag and 3.589$\pm$0.023\,mag,
respectively). 
There are only three discarded candidate systems, WDS~10197+1928, 
WDS~11455+4740, and WDS~18111+3241, with the (incorrect) WDS note about binarity
``V'' (``proper motion or other technique indicates that this pair is 
physical''). 

The proper motions of the faint white dwarfs WD~0223--092 and WD~0221--095 in
the WDS~02255--0904 system differ by about 14\,mas\,a$^{-1}$, which translates
into a relative difference of about 17\,\%. 
Given the relatively low absolute value of the proper motions (of the
same order of those of typical background thick disc and halo stars and white
dwarfs), the very large expected projected physical separation ($s \sim$ 0.6\,pc
for a minimum heliocentric distance of $d$ = 100\,pc), and the low mass of the
objects ($M_A \sim M_B \lesssim$ 1\,$M_\odot$), asserting that the system may be
gravitationally bound is rather speculative. 

   \begin{table}
      \caption[]{New proper motions of components in wide binary candidates.} 
         \label{newpropermotions}
     $$ 
         \begin{tabular}{ll cc c}
            \hline
            \hline
            \noalign{\smallskip}
WDS  	        & Name		& $\mu_\alpha \cos{\delta}$	& $\mu_\delta$          & $\Delta \mu$  	\\
identifier      &		& [mas\,a$^{-1}$]		& [mas\,a$^{-1}$]       & [mas\,a$^{-1}$]	\\
            \noalign{\smallskip}
            \hline
            \noalign{\smallskip}
00059+1805	& HD 101$^a$	& --152.2$\pm$1.1		& --148.1$\pm$1.4	& 9.1$\pm$1.7  		\\ %
		& LP 404--21	& --146.6$\pm$0.4		& --140.9$\pm$0.4	&	  		\\ %
01024+0504	& HD 6101 AB$^a$& +323.3$\pm$1.2	       	& +226.0$\pm$1.2       	& 6$\pm$2  		\\ %
		& G~1--45 AB	& +329.3$\pm$0.5	       	& +223.7$\pm$1.0       	& 		  	\\ %
02255--0904	& WD~0223--092	& +82.0$\pm$1.5			& +11.3$\pm$1.0		& 14$\pm$3  		\\ %
		& WD~0221--095	& +82.8$\pm$1.9			& --2.9$\pm$1.7		& 		  	\\ %
02310+0823	& G 73--63$^a$	& +376.1$\pm$1.9	       	& --85.4$\pm$1.6       	& 50$\pm$3  		\\ %
		& G 73--59	& +344.7$\pm$0.9	       	& --124.2$\pm$1.5       & 		  	\\ %
03162+5810	& GJ 130.1 A$^a$& +445.6$\pm$3.9		& --340.3$\pm$4.1	& 92$\pm$6  		\\ %
		& G 246--30	& +479.4$\pm$0.7		& --255.1$\pm$1.0	& 		  	\\ %
10197+1928	& 40~Leo$^a$	& --230.2$\pm$0.6	       	& --214.6$\pm$0.4       & 25$\pm$2  		\\ %
		& LP 371--59~A$^b$& --223.5$\pm$1.3	       	& --238.2$\pm$1.6      	&  			\\ %
11452+1821	& G 57--17	& --297.1$\pm$1.9		& --291.3$\pm$1.2	& 50$\pm$3  		\\ %
		& G 57--15	& --258.0$\pm$1.4		& --260.9$\pm$0.7	& 		  	\\ %
11455+1821	& HD 102158$^a$	& --591.6$\pm$0.7		& --290.7$\pm$0.5	& 91.6$\pm$1.5  	\\ %
		& G 122--46	& --581.2$\pm$1.0		& --199.7$\pm$0.7	& 		  	\\ %
16348--0412	& HD 149414 AB$^a$& --133.7$\pm$1.4		& --701.2$\pm$1.4	& 61$\pm$2  		\\ %
		& BD--03 3968B$^c$& --191.2$\pm$0.7		& --680.0$\pm$1.0	& 		  	\\ %
18111+3241	& BD+32 3065$^a$& --134.7$\pm$1.2		& +322.1$\pm$1.3	& 28$\pm$2  		\\ %
		& G 206--16	& --161.1$\pm$0.8		& +326.8$\pm$1.3	& 6$\pm$2$^d$  		\\ %
		& NLTT 46103	& --156.7$\pm$0.5		& +322.4$\pm$1.6	& 		  	\\ %
22175+2335	& G 127--13	& --93.1$\pm$1.1		& --383.7$\pm$0.4	& 36.4$\pm$1.5  	\\ %
		& G 127--14	& --115.7$\pm$0.7		& --412.2$\pm$0.5	& 		  	\\ %
23228+2208	& BD+21 4923$^a$& +198.3$\pm$1.2		& --69.8$\pm$1.2	& 94$\pm$3  		\\ %
		& G 68--7	& +276.1$\pm$1.8		& --122.1$\pm$1.3	& 		  	\\ %
            \hline
         \end{tabular}
     $$ 
\begin{list}{}{}
\item[$^{a}$] Proper motions of bright primaries are from R\"oser et~al.~(2008).
\item[$^{b}$] LP~371--59~A is the primary in a close binary system (see text).
\item[$^{c}$] Bakos et~al. (2002) tabulated ``revised proper motions'' for 
BD--03~3968B that were wrong by almost 1000\,mas\,a$^{-1}$.
\item[$^{d}$] $\Delta \mu$ between G~206--16 and NLTT~46103.
\end{list}
   \end{table}

Of the other {two} systems, the identical parallactic distances and similar
proper motions and isochronal ages of binary star HD~6101~AB and white dwarf
G~1--45~AB in system WDS~01024+0504 support a true physical connection
(see below). 
However, HD~101 and LP~404--21 in system WDS~00059+1805, although having 
similar proper motions, are {\em not} physically connected.
WDS~00059+1805 is a hierarchical triple system at $d$ = 37.1$\pm$0.8\,pc.
The non-tabulated system members are \object{HD~113}~A and~B (HIP~495), which is
a K0+K0 binary ($\rho$ = 3.445$\pm$0.004\,arcsec, $\Delta H_P$ =
0.29$\pm$0.02\,mag) at about 9.4\,arcmin to the south of the F8 primary HD~101. 
The hypothetical fourth component, LP~404--21, is 7.7\,mag fainter in the $J$
band than the primary. 
This magnitude difference would imply that LP~404--21 is an M5--6 dwarf with a
colour $V-J \sim$ 5.5\,mag if it were located at the same heliocentric distance
to HD~101 (assuming a typical age of 0.5--5\,Ga).
However, its actual $V-J$ colour is only about 2.5\,mag and LP~404--21 is thus
a late-K- or early-M-type dwarf or subdwarf at a larger heliocentric distance.
L\'epine \& Bongiorno (2007) also suggested its subdwarf nature.

\subsection{Probable bound systems}
\label{probableboundsystems}

From previous section, there remains only {seven} systems ({six} with
reliable common proper motions from {\em Hipparcos} or Tycho-2 catalogues plus
the HD~6101~AB + G~1--45~AB system) with high probability of being physically
connected. 
Next, I~discuss them in detail.

\subsubsection{WDS~01024+0504 (HD~6101~AB and G~1--45~AB)}

This is a hierarchical quadruple system.
HD~6101~AB is a relatively bright ($V$ = 8.16\,mag) close binary star of
combined spectral type K3V and low activity ($\log{R'_{\rm HK}}$ = --4.661; Gray
et~al. 2003).
It was first resolved by the {\em Hipparcos}  mission ($\rho$ =
0.711$\pm$0.011\,arcsec, $\Delta H_P$ = 1.87$\pm$0.04\,mag). 
Afterwards, it has been astrometrically followed up by several authors (Mason
et~al. 1999; Balega et~al. 2002, 2004, 2007; Richichi et~al. 2007).
Using mostly speckle interferometric observations, Balega et~al. (2006)
presented new orbital parameters for HD~6101~AB, from where they derived a
period $P$ = 29.0$\pm$0.6\,a, a semi-major axis $a$ = 9.8$\pm$0.3\,AU, and a
total mass $\mathcal{M_{\rm A} + M_{\rm B}}$ = 1.17$\pm$0.14\,$M_\odot$.
Accounting for the magnitude difference in the optical ($\Delta V \approx$
1.7\,mag), the Siess et~al. (2000) grid of tracks for low- and intermediate-mass
stars of 1--5\,Ga, and the Balega et~al. (2006) total mass, one may derive that
the secondary must have a spectral type between K7V and~M2V.  

At 1276\,arcsec to the east of HD~6101~AB, it is located the binary white dwarf
G~1--45~AB (WD~0101+048). 
It was discovered in the Lowell proper motion survey of Giclas et~al. (1959),
who assigned it a DAs spectral type (currently, it is determined at DA5). 
Because of its relative brighness ($V$ = 14.10\,mag), G~1--45~AB has been
investigated and catalogued in numerous occasions (Shipman 1979; Green et~al.
1986; Liebert et~al. 1988; McCook \& Sion 1999; Bergeron et~al. 2001; Zuckerman
et~al. 2003; Farihi et~al. 2005; Mullally et~al. 2007).
The white dwarf is a double degenerate, as it shows radial velocity variations
with an uncertain period of 0.7--6.5\,d (Saffer et~al. 1998; Maxted et~al.
2000).  
The spectroscopic (total) mass of 0.77\,$M_\odot$ provided by Lajoie \& Bergeron
(2007) is consistent with the parallactic distance of G~1--45~AB measured by
van~Altena et~al. (1995) and of HD~6101~AB measured by {\em Hipparcos} (however,
many papers list photometric distances at about 13.5\,pc, which do {\em not} fit
white dwarf theoretical models). 

Proper motions of both binary objects differ by only 6$\pm$3\,mas\,a$^{-1}$
(Section~\ref{dedicatedfollowup}).
Because of the similarity in parallactic distance and proper motion, the WDS
note ``V'' about the wide binarity of HD~6101~AB and G~1--45~AB 
(``[...] this pair is physical'') may be correct.

\subsubsection{WDS~13090+3353 (LP~268--35 and LP~268--33)}

The system LEP~62AC was proposed by L\'epine \& Bongiorno (2007).
It is, therefore, one of the very wide binary candidates in this work that have
been identified more recently. 
It is also the faintest system in this section (only the primary is listed in
the {\em Hipparcos} catalogue).
As a result, both stars have been poorly investigated.
The most remarkable fact in the literature is that Ryan (1992) classified the
primary, LP~268--35, as a normal dwarf based on $UBVRI$ photometry (i.e. it is
not a subdwarf of the Galactic halo). 
Previously, it had been proposed in one of the Luyten proper motion catalogues
that the primary forms a closer pair with a star located at about 3\,arcmin to
the southwest (\object{LP~268--34}).
The USNO-B1 proper motion of this hypothetical companion, ``WDS~13090+3353~B'',
is ($\mu_\alpha \cos{\delta}$, $\mu_\delta$) $\approx$ (--218,
--38)\,mas\,a$^{-1}$, consistent with an accurate measurement by L\'epine \&
Shara (2005), but different from those of LP~268--35 and LP~268--33 by more than
70\,mas\,a$^{-1}$ (i.e. LP~268-34 is not part of the proper motion
system).

In Table~\ref{wds13090+3353}, I~compile SDSS and 2MASS photometry of both
LP~268--35 and LP~268--33. 
There is a good agreement between the observed magnitudes and those expected for
K7--M1V and M4--5V stars at the {\em Hipparcos} distance of the primary, $d$ =
66$\pm$12\,pc. 
For the comparison, I~have used the colours and absolute magnitudes as functions
of late spectral type as tabulated by Bochanski et~al. (2007), West et~al.
(2008), and Caballero et~al. (2008).
While the fit of the colours from $rizJHK_{\rm s}$-band magnitudes of LP~268--33
to an M4.5$\pm$0.5V template is excellent, the dwarf displays an obvious blueing
in the $u$ and $g$ bands. 
This is not unforeseen, since roughly 50\,\% of M4--5V stars display activity,
which is associated to an excess of flux in the blue optical (West et~al. 2008).
Activity lifetimes of M4--5V stars vary between 4.0 and 7.5\,Ga, which may be
understood as a (conservative) upper limit for the age of LP~268--33.

The difference in proper motion between LP~268--35 and LP~268--33, 
($\Delta \mu_\alpha \cos{\delta}$, $\Delta \mu_\delta$) = 
(1$\pm$6, 11$\pm$6)\,mas/a, is null within 1--2$\sigma$.
Given the resemblance between proper motions and parallactic and photometric
distances of the two of them, I~will assume that they {\em travel together}
through the Galaxy.
No radial velocity measurements exist for the two dwarfs, from where one could
confirm the common space velocity or identify membership in a young moving
group.

Using the 2MASS $J$-band magnitudes in Table~\ref{wds13090+3353}, the
hypothetical common distance $d$ = 66$\pm$15\,pc, a solar age, and the
theoretical models of Baraffe et~al. (1998), I~derive masses $\mathcal{M}_{\rm
A}$ = 0.71$\pm$0.08\,$M_\odot$ and $\mathcal{M}_{\rm B}$ =
0.32$\pm$0.09\,$M_\odot$ for LP~268--35 and LP~268--33 (at late spectral types,
the near-infrared $J$ band works better for comparison with the Lyon theoretical
models than optical ones -- e.g.~$V$).  

   \begin{table}
      \caption[]{Photometry of system WDS~13090+3353.} 
         \label{wds13090+3353}
     $$ 
         \begin{tabular}{l cc}
            \hline
            \hline
            \noalign{\smallskip}
Magnitude  		& LP~268--35		&  LP~268--33		\\
            \noalign{\smallskip}
            \hline
            \noalign{\smallskip}
$u$ [mag]		& 15.480$\pm$0.007	& 19.860$\pm$0.038	\\ %
$g$ [mag]		& 12.588$\pm$0.001	& 17.175$\pm$0.007	\\ %
$r$ [mag]		& 11.126$\pm$0.001	& 15.727$\pm$0.006	\\ %
$i$ [mag]		& 10.720$\pm$0.001	& 14.135$\pm$0.006	\\ %
$z$ [mag]		& 10.887$\pm$0.002$^{a}$& 13.265$\pm$0.007	\\ %
$J$ [mag]		&  9.296$\pm$0.019	& 11.685$\pm$0.021	\\ %
$H$ [mag]		&  8.740$\pm$0.018	& 11.052$\pm$0.021	\\ %
$K_{\rm s}$ [mag]	&  8.623$\pm$0.018	& 10.769$\pm$0.019	\\ %
            \hline
         \end{tabular}
     $$ 
\begin{list}{}{}
\item[$^{a}$] The $z$-band measurement of LP~268--35 is probably affected by
non-linearity of the detector.
\end{list}
   \end{table}

\subsubsection{WDS~14396--6050 ($\alpha$~Cen~AB and Proxima)}

The $\alpha$~Cen system has been subject of intensive and extensive studies in
the literature (Gasteyer 1966; Kamper \& Wesselink 1978; Matthews
\& Gilmore 1993; Wertheimer \& Laughlin 2006).
As already mentioned in Section~\ref{introduction}, Anosova \& Orlov (1991)
proposed that $\alpha$~Cen~AB and Proxima are stars in the ``moving group
of $\alpha$~Cen''.
The other stars in the moving group would be the binary \object{HD~21209}~AB
(K3.5V + K8Vk:) and the triple system \object{V1089~Her} and
\object{V1090~Her}~AB (K5.0V + [K5.0V + M1.0V]; Reid et~al. 2004). 
Anosova et~al. (1994) went on the discussion\footnote{A critical reading is
needed: Anosova et~al. (1994) assumed a mass of 0.020\,$M_\odot$ for Proxima,
close to the brown dwarf-planet boundary, which is about five times lower than
currently assumed.}, and enlarged the list of ``satellites of $\alpha$~Cen''.
However, in spite of the chromospheric activity of Proxima (with flares and
strong Mg~{\sc ii}~h+k $\lambda$280\,nm in emission), the $\alpha$~Cen triple
system is accepted to be relatively old, with an age at about 5--6\,Ga.  
Since I~do not present new data that help answering the original question in
Vo\^ute (1917), that if {``they [$\alpha$~Cen~AB and Proxima] are physically
connected or members of the same drift''}, I~follow Ludwig Wittgenstein's
proposition of {``passing over in silence''} and will follow the general
agreement that they {\em are} gravitationally bound.

The masses compiled by Wertheimer \& Laughlin (2006) for $\alpha$~Cen~AB and
Proxima were 2.039$\pm$0.009\,$M_\odot$ (combined) and 0.11$\pm$0.02\,$M_\odot$,
respectively.

\subsubsection{WDS~15208+3129 (HD~136654 and BD+32~2572)}

This system, formed by an F5V and a K0V star, was also proposed by L\'epine \&
Bongiorno (2007). 
However, in contrast to system WDS~13090+3353, there exist {\em Hipparcos}
parallax measurements for both HD~136654 and BD+32~2572.
From the new data reduction by van~Leeuween (2007), the differences between
proper motions and parallactic distances are 
($\Delta \mu_\alpha \cos{\delta}$, $\Delta \mu_\delta$) = 
(1.3$\pm$0.7, 0.6$\pm$1.1)\,mas/a and $\Delta d$ = 1$\pm$2\,pc.
The difference between radial velocities is also very small and probably
not significant: $\Delta V_r$ = 0.7$\pm$0.5\,km\,s$^{-1}$ (Montes et~al. 2001;
Nordstr\"om et~al. 2004).
As a result, they seem to form a real common proper motion~pair.

The primary in the system, HD~136654, is a single (Mason et~al. 2001),
non-variable (McMillan et~al. 1976), high metallicity (Fischer \& Valenti
2005; Robinson et~al. 2006) star. 
The secondary, BD+32~2572, has not been so well investigated:
Strassmeier et~al. (2000) found Ca~{\sc ii} H+K in emission
and Violat-Bordonau \& Violat-Mart\'{\i}n (2006) measured a low amplitude of
photometric variability ($\Delta V \approx$ 0.18\,mag) with a period near
9.24\,d. 
However, the most important fact is that BD+32~2572 is a probable member in the
\object{Hyades Supercluster} (``Kapteyn's stream~I''; Montes et~al. 2001), which
automatically makes HD~136654 to be in as well. 
Indeed, the HD~136654 high metallicity and space velocities from Nordstr\"om
et~al. (2004) and  Karata{\c s} et~al. (2004) agree well with such a membership
and, thus, an Hyades-like age.
At $\tau \sim$ 600\,Ma, theoretical masses of HD~136654 and BD+32~2572 are
$\mathcal{M}_{\rm A} \sim$ 1.2--1.3\,$M_\odot$ and $\mathcal{M}_{\rm B} \sim$
0.9--1.0\,$M_\odot$ (Baraffe et~al. 1998; Siess et~al.~2000).

\subsubsection{WDS 20124--1237 ($\xi^{02}$~Cap and LP~754--50)}

The primary star, $\xi^{02}$~Cap, is a single (McAlister et~al. 1987;
Lagrange et~al. 2009), F7V-type star that has been subject of numerous analyses.
Some basic stellar parameters are: $T_{\rm eff} \approx$ 6330\,K, 
[Fe/H] $\approx$ --0.27, $\mathcal{M} \approx$ 1.10\,$M_\odot$, 
$\log{\epsilon_{\rm Li}} \approx$  2.79, $\tau \approx$ 4.78\,Ga 
(Chen et~al. 2001; Lambert \& Reddy 2004). 
Its solar age is consistent with a relative large modulus of vertical
heliocentric space velocity component, W = --42\,km\,s$^{-1}$ (Nordstr\"om
et~al. 2004).

The secondary star, LP~754--50, has been referenced few times in the literature.
It is a Luyten star whose astrometry was improved by Salim \& Gould (2003) and
whose spectral type was determined at M0Vk by Gray et~al. (2006).
The latter authors also measured $\log{R'_{\rm HK}}$ = --4.699, at the
active-inactive boundary (but with redder $B-V$ colour).
The star was tabulated in the {\em Hipparcos} catalogue.
Its parallactic distance in the new reduction by van~Leeuwen (2007), $d$ =
23.6$\pm$1.6\,pc, differs from that in the original catalogue by Perryman et~al.
(1997), $d$ = 26.4$\pm$1.9\,pc, and from those of $\xi^{02}$~Cap ($d$ =
27.7$\pm$0.3 and 28.1$\pm$0.7\,pc, respectively).
In general, the new reduction by van~Leeuwen (2007) provided better accuracies,
by up to a factor four, than in the original {\em Hipparcos} catalogue.
However, it was not infallible.
For example, Caballero \& Dinis (2008) showed some stars whose astrometric
solution got worse with the new reduction.
If LP~754--50 were located at the distance to $\xi^{02}$~Cap, it would have an
absolute $J$-band magnitude $M_J$ = 6.27$\pm$0.03\,mag, which translates into 
a theoretical mass $\mathcal{M} \sim$ 0.55\,$M_\odot$ and an effective
temperature typical of an early M dwarf of solar age (Baraffe et~al. 1998).
Using the original {\em Hipparcos} astrometry (parallaxes, proper motions),
there are grounds to consider WDS~20124--1237 a physical pair.

\subsubsection{WDS 20452--3120 (AU~Mic and AT~Mic~AB)}

It is a late-type, hierarchical triple system in the very young $\beta$~Pictoris
moving group ($\tau \sim$ 12\,Ma).
The secondary, AT~Mic~AB, is a binary resolved by {\em Hipparcos} 
($\rho$ = 3.349$\pm$0.007\,arcsec, $\Delta H_P$ =
0.09$\pm$0.06\,mag). 
Numerous investigations and reviews have targeted the {\em three} stars, which
are among the youngest pre-main sequence stars in the solar neighbourhood.
They display X-ray emission, flaring activity, flux excess in the infrared
due to a circumstellar disc (AU~Mic), photometric variability (of BY~Dra type),
active coronae, Ca~{\sc ii} H+K in emission and other properties typical in
young stars  
(Linsky et~al. 1982; Kundu et~al. 1987; Pallavicini et~al. 1990; Batalha et~al.
1996; Barrado y Navascu\'es et~al. 1999; Katsova et~al. 1999; Zuckerman et~al.
2001; Kalas et~al.~2004).  
There is a difference between {\em Hipparcos} heliocentric distances of
0.8$\pm$0.4\,pc, which could be real (as in the case of $\alpha$~Cen~AB and
Proxima) or be due to a poor accuracy in the parallax measurement (e.g. as in
the case of WDS~13090+3353 and WDS~20124--1237). 

Using the parallactic distances by van~Leeuwen (2007), the 2MASS $JHK_{\rm
s}$-band magnitudes, an estimated age $\tau$ = 12$^{+8}_{-4}$\,Ma, and the
NextGen models from Baraffe et~al. (1998), and assuming that the difference of
magnitudes in the near infrared between AT~Mic~A and AT~Mic~B are $\Delta
(JHK_{\rm s}) \lesssim \Delta H_P$, I~derive masses of 0.45$\pm$0.10,
0.27$^{+0.04}_{-0.09}$, and 0.25$^{+0.04}_{-0.09}$\,$M_\odot$ for AU~Mic (M1Ve),
AT~Mic~A (M4.5Ve), and AT~Mic~B (M5:). 

Although the stars in the $\beta$~Pictoris moving group are spread over a
space region with a size of only about 74\,pc (Ortega et~al. 2002), the
short separation between AU~Mic and AT~Mic~AB (projected physical separation $s$
= 0.226$\pm$0.002\,pc) is remarkable and may indicate a common origin within the
birthplace.

   \begin{table*}
      \caption[]{Probable bound wide systems.} 
         \label{probableboundwidesystems}
     $$ 
         \begin{tabular}{lll ccccc}
            \hline
            \hline
            \noalign{\smallskip}
WDS  	        & Primary	& Secondary	& $s$		&$\mathcal{M}_1$&$\mathcal{M}_2$& $P^*$		& $U_g^*$ 	\\
identifier      &		&		& [10$^3$\,AU]  & [M$_\odot$]	& [M$_\odot$]	& [Ma]		& [10$^{33}$\,J]\\
            \noalign{\smallskip}
            \hline
            \noalign{\smallskip}
01024+0504	& HD~6101~AB	& G~1--45~AB	& 26.9$\pm$0.6	& 1.17  	& 0.77	  	& 3.2		& --59.1	\\ %
13090+3353	& LP 268--35	& LP 268--33	& 84$\pm$15 	& $\sim$0.7   	& $\sim$0.3	& 24		& --4.4    	\\ %
14396--6050	&$\alpha$ Cen AB& Proxima	&12.0$\pm$0.6$^{a}$& 2.039   	& 0.11	     	& 0.90$^{a}$	& --32.1$^{a}$  \\ %
15208+3129	& HD 136654	& BD+32 2572	& 68.8$\pm$1.7 	& $\sim$1.2	& $\sim$0.9	& 12		& --28		\\ %
20124--1237	& $\xi^{02}$ Cap& LP 754--50	& 28.3$\pm$0.3 	& 1.10   	& 0.55	     	& 3.7		& --37.8    	\\ %
20452--3120	& AU Mic	& AT Mic AB	& 46.4$\pm$0.5	& 0.45   	& 0.52	     	& 10		& --7.7    	\\ %
20599+4016	& HD 200077 AE--D& G 210--44 AB	& 49.7$\pm$1.1 & $\sim$2.9   	& $\sim$1.2   	& 5.5		& --120    	\\ %
            \hline
         \end{tabular}
     $$ 
\begin{list}{}{}
\item[$^{a}$] True physical separation, orbital period, and gravitational energy
accounting for the different heliocentric distances. 
\end{list}
   \end{table*}

\subsubsection{WDS 20599+4016 (HD~200077~AE--D and G~210--44~AB)}

It is a hierarchical quintuple system of complicated nomenclature and structure.
On the one hand, HD~200077 is a triple star of combined F8V spectral type.
In the year 1908, Burnham proposed two stars of $V \sim$ 6--10\,mag at $\rho
\sim$  2--3\,arcmin to the southeast and southwest of HD~200077, labelled ``B''
and ``C'', to be common proper motion companions.
They are, however, background stars of lower proper motion.
Almost a century later, the {\em Hipparcos} mission resolved HD~200077 into a
close binary with  $\rho$ = 1.949$\pm$0.027\,arcsec, $\Delta H_P$ =
4.07$\pm$0.10\,mag;  
the faintest component receives the label ``D''.
The brightest one is, in its turn, a double-lined spectroscopic binary
discovered by Latham et~al. (1988) and confirmed by Goldberg et~al. (2002);  
the low-mass spectroscopic companion is labelled ``E''.
Mazeh et~al. (2003) tabulated $P_{\rm AE}$ = 112.55$\pm$0.04\,d, $q_{\rm AE}$ =
0.85$\pm$0.02, and [Fe/H]$_{\rm AE}$ = --0.40.
These authors used the estimated mass $\mathcal{M}_{\rm A} \sim$ 0.84\,$M_\odot$
for the primary from Carney et~al. (1994), which is inconsistent with the F8V
spectral type and several effective temperature determinations of components A
and E (e.g. Goldberg et~al. 2002).
These determinations favour masses $\mathcal{M}_{\rm A} \sim$
1.1--1.3\,$M_\odot$ and $\mathcal{M}_{\rm E} \sim$ 0.9--1.0\,$M_\odot$ (and
spectral types G1V and G6--9V, respectively). 
The component D, given its magnitude difference with respect to AE, may be
a late K-type dwarf with a mass $\mathcal{M}_{\rm D} \sim$ 0.7\,$M_\odot$.
The triple system as a whole does not display X-ray activity (Ottmann
et~al.~1997). 

On the other hand, G~210--44~AB, of combined spectral type M1V, is another close
binary first resolved by {\em Hipparcos} ($\rho$ = 0.334$\pm$0.023\,arcsec,
$\Delta H_P$ = 1.30$\pm$0.27\,mag -- see also Balega et~al. 2004, 2007).
Using the theoretical models of Baraffe et~al. (1998), G~210--44~AB matches the
scenario of an 0.65\,$M_\odot$- and an 0.55\,$M_\odot$-mass pair at $d$ =
41.0$\pm$0.9\,pc moving in the same direction as HD~200077~AE--D.

\section{Discussion}
\label{discussion}

\subsection{The 0.1\,pc ``cutoff'', young moving groups, orbital periods, and
missing binaries}

Parallax, proper motion, and photometry measurements are all consistent with
the seven systems being physical pairs.
Six of them have projected physical separations $s >$ 0.1\,pc
($s >$ 2\,10$^4$\,AU; fourth column in Table~\ref{probableboundwidesystems}).
The exception is the $\alpha$~Cen system.
Even accounting for the different heliocentric distances, the true physical
separation between $\alpha$~Cen~AB and Proxima, $r$ = 0.058$\pm$0.003\,pc ($r >
s$), is shorter than the tenth of a parsec. 
However, the existence of six systems with $s >$ 0.1\,pc, if really bound, shows
that there are deviations to the hypothetical cutoff in binary frequency at this
value, as proposed originally by Bahcall \& Soneira (1981) and Retterer \& King
(1982). 

The widest systems are WDS~13090+3353 (LP~268--35; $s$ = 0.41$\pm$0.07\,pc) and
WDS~15208+3129 (HD~136654; $s$ = 0.334$\pm$0.008\,pc), which is a probable
member in the Hyades Supercluster ($\tau \sim$ 600\,Ma).
Systems WDS~20599+4016 (HD~200077) and AU~Mic in the $\beta$~Pictoris group
($\tau \sim$ 12\,Ma) have also projected physical separations larger than
0.2\,pc. 
Three of the latter four systems were first proposed by L\'epine \& Bongiorno
(2007) as faint companions of {\em Hipparcos} stars.
This is an indication of how much we must still learn of wide binarity in the
solar neighbourhood.

Remarkably, two of the seven systems belong to young kinematic groups (HD~136654
and AU~Mic). 
This is in agreement with the simple idea that young wide binary systems have
had less time to be perturbed and disrupted by Galactic material of all types.
However, a few of the remaining systems have relatively well determined ages at
about the Solar value, such as WDS~01024+0504 (HD~6101, which has a relatively
cool binary white dwarf companion that had to leave the main sequence several
10$^8$\,a ago), $\alpha$~Cen, and WDS 20124--1237 ($\xi^{02}$~Cap).
The absence of X-ray emission from HD~200077 probably indicates that its system
is older than 1\,Ga as well.
It leaves only the LP~268--35 system (the widest one) as a suitable target to
investigate its membership in a young kinematic group.

The minimum orbital periods $P^*$ (Table~\ref{probableboundwidesystems}) range
between less than 1\,Ma for $\alpha$~Cen and 24\,Ma for LP~268--35.
I~computed $P^*$ using the Kepler's Third Law and the projected physical
separation between components in system, $s$, assuming that they are located at
the distance of the primary (i.e. $s$ instead of $r$ or $a$). 
With an age not older than $\tau \sim$ 20\,Ma, the AU~Mic system in the
$\beta$~Pictoris moving group has completed only two orbital periods at 
most since its formation.
However, the HD~136654 system in the Hyades Supercluster ($\tau \sim$ 600\,Ma),
although it is also young and has a long orbital period of about 12\,Ma, has
revolved roughly 50 times about a common centre of mass.
The rest of the wide systems have had time enough to complete several hundred
orbits.  

There can be missing binaries in the WDS catalogue with angular
separations $\rho <$ 1000\,arcsec but projected physical separations $s >$
10$^5$\,AU (i.e. at larger heliocentric distances than the systems studied here
and, therefore, more difficult to follow-up in general) or even not listed in
the WDS catalogue. 
For example, the system \object{Fomalhaut} + \object{TW~PsA} in the young Castor
moving group was proposed by Gliese (1969) and does not appear in the catalogue
as a possible wide binary ($\rho \approx$ 7100\,arcsec, $r$ = 0.28$\pm$0.03\,pc,
$\Delta \mu$ = 6.1$\pm$0.8\,mas\,a$^{-1}$).
The halo system HD~149414 (Section~\ref{introduction}) is neither in the WDS.
The search for such missing wide binaries will be carried out in another work.

\subsection{A comparison of binding energies}  

\begin{figure}
\centering
\includegraphics[width=0.49\textwidth]{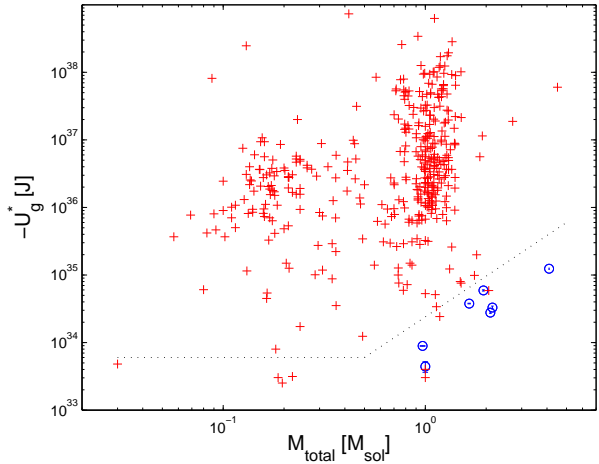}
\caption{Binding energy-total mass diagram.
Open (blue) circles with error bars are for the seven systems in
Section~\ref{probableboundsystems} and (red) crosses are for the ensamble of
systems with very low-mass components presented in the text.
The dotted line indicates a boundary for the selection of multiple systems
in the solar neighbourhood with at least one low-mass component ($\mathcal{M} <$
0.15\,$M_\odot$) and the lowest binding energies for their total masses.}   
\label{xmUgvsMtotal}
\end{figure}
%

   \begin{table*}
      \caption[]{Multiple systems in the solar neighbourhood with at least one
      low-mass component ($\mathcal{M} <$  0.15\,$M_\odot$) and the lowest
      binding energies for their total masses.} 
         \label{knownfragile}
     $$ 
         \begin{tabular}{lll cccc l}
            \hline
            \hline
            \noalign{\smallskip}
WDS  	        	& Primary	& Secondary	& $s$		&$\mathcal{M}_1$&$\mathcal{M}_2$& $U_g^*$ 	& Discovery	\\
identifier      	&		&		& [10$^3$\,AU]  & [M$_\odot$]	& [M$_\odot$]	& [10$^{33}$\,J]& reference \\
            \noalign{\smallskip}
            \hline
            \noalign{\smallskip}
00212--4246$^{a}$	& K\"o~1~A	& K\"o~1~B	& 1.8		& 0.103  	& 0.079		& --8.0		& Caballero 2007a \\ %
01269--5023		& 2M0126--50~A	& 2M0126--50~B	& 5.1		& 0.095  	& 0.092		& --3.0		& Artigau et~al. 2007 \\ %
12076--3933		& 2M1207--39~A	& 2M1207--39~B	& 0.046		& 0.025  	& 0.005		& --4.8		& Chauvin et~al. 2004 \\ %
...$^{b}$		& 2M1258+40~A	& 2M1258+40~B	& 6.7		& 0.105  	& 0.091		& --2.5		& Radigan et~al. 2009 \\ %
15232+3017$^{c}$	& $\eta$~CrB~AB	& $\eta$~CrB~C	&  3.6		& 2.000  	& 0.060		& --59		& Kirkpatrick et~al. 2001 \\ %
23315--0405$^{d}$	& K\"o~3~A	& K\"o~3~BC	& 11.9		& 1.02  	& 0.160		& --24		& Caballero 2007b \\ %
...$^{e}$		& Sun		& Uranus	& 0.019191	& 1.000  	&4.367\,10$^{-5}$& --4.0	& Herschel 1783 \\ %
...$^{f}$		& Sun		& Neptune	& 0.030069	& 1.000  	&5.151\,10$^{-5}$& --3.0	& Le Verrier 1847 \\ %
            \hline
         \end{tabular}
     $$ 
\begin{list}{}{}
\item[$^{a}$] The (abridged) names of the components in the Koenigstuhl~1 
system are \object{LEHPM~494} (K\"o~1~A) and DE0021--42 (K\"o~1~B).
\item[$^{b}$] The system 2M1258+40~AB awaits a WDS numbering.
\item[$^{c}$] Kirkpatrick et~al. (2001) used the name ``Gl~584C'' for the brown 
dwarf companion $\eta$~CrB~C. 
The primary $\eta$~CrB~AB is a spectroscopic binary with individual masses 1.003 
and 0.997\,M$_\odot$.
\item[$^{d}$] The (abridged) names of the components in the Koenigstuhl~3 
system are \object{HD~221356} (K\"o~3~A) and 2M2331--04~AB (K\"o~3~BC).
The secondary is, in its turn, a M8.0V~+~L3.0V close binary with individual
masses 0.088 and 0.072\,M$_\odot$ (Gizis et~al. 2000, 2003; Caballero 2007b). 
\item[$^{e}$] Actually, Uranus was pointed out by sir William Herschel in 
March~1781. 
\item[$^{f}$] There is a consensus that Urbain Le~Verrier, John Couchh Adams, 
and Johann Galle jointly deserve credit for discovering Neptune. 
\end{list}
   \end{table*}

Wide binarity is synonymous with multiplicity of systems with very low (absolute
values of) gravitational potential energies, $U_g = -G M_1 M_2 / r$. 
In the last column of Table~\ref{probableboundwidesystems}, I~show the 
gravitational potential (binding) energy, $U_g^*$, using $r \sim s$ (except for 
$\alpha$~Cen), and the corresponding combined masses (e.g. at a large
separation, Proxima feels the gravitational attraction of $\alpha$~Cen~AB as if 
it were a single, more massive star). 
The asterisk in $U_g^*$ indicates that the absolute values of the ``true''
potential energies $U_g$ using the physical separation $r$ must be lower than in
Table~\ref{probableboundwidesystems}.

In Fig.~\ref{xmUgvsMtotal}, I~show a $-U_g^*$ vs. $M_1 + M_2$ diagram for the
seven systems in Section~\ref{probableboundsystems} (circles) and a collection
of 399 multiple systems including mostly:  
\begin{itemize}
\item the Sun and the four giant planets in the Solar System (Jupiter, Saturn,
Uranus, and Neptune); 
\item transit and radial-velocity exoplanets and candidates from the Extrasolar
Planet Encyclopaedia;
\item ``classic'' late-M-type binaries of the Solar neighbourhood with mass
ratios $q > 0.5$, such us \object{EZ~Aqr}~AB, \object{EI~Cnc}~AB,
\object{QY~Aur}~AB, and \object{GJ~1005}~AB;
\item field late-M-, L-, and T-type binaries in systems with $q > 0.5$ (e.g.
Lane et~al. 2001; Bouy et~al. 2003; Siegler et~al. 2005; Forveille et~al. 2005;
Burgasser \& McElwain 2006; Caballero 2007a -- see also Bouy et~al. 2005 and
Burgasser et al. 2007 for compilations);
\item ``classic'' systems in the Solar neighbourhood with late-M-type companions
and $q < 0.5$, such us \object{V1054~Oph} + \object{GJ~643} +
\object{vB~8}, \object{GX~And} + \object{GQ~And}, \object{EQ~Peg}~AB,
\object{V1428~Aql} + \object{vB~10}, and \object{$o^2$~Eri}~AC;
\item field stellar systems with late-M-, L-, and T-type companions and $q <
0.5$ (e.g. Rebolo et~al. 1998; Goldman et~al. 1999; Burgasser et~al. 2000;
Kirkpatrick et~al. 2001; Gizis et~al. 2001; Scholz et~al. 2003; Seifahrt et~al.
2005). 
\end{itemize}
This sample was used by Caballero (2007b) for comparison purposes and
was quite complete for systems with at least one planet, brown dwarf, or
low-mass star with $M_2 \lesssim$ 0.15\,$M_\odot$ (i.e. with low $M_1 M_2$
product or, inversely, small expected $|U_g^*|$).
Afterwards, it has been updated with new discoveries (e.g. Radigan et~al. 2009).

The seven systems in Section~\ref{probableboundsystems} are among the
multiple systems in the solar neighbourhood with the lowest binding energies for
their total masses.
Of the collection of 399 multiple systems, only eight (not counting
$\alpha$~Cen) have comparable low values of $|U_g^*|$
(Table~\ref{knownfragile}; they are the systems below the dotted line in 
Fig.~\ref{xmUgvsMtotal}).  
Two of them are not resolved multiple stellar systems, but planetary
systems: the Sun and Uranus, and the Sun and Neptune.

The brown dwarf-exoplanet pair 2M1207--39~AB (Chauvin et~al. 2004, 2005),
because of its large mass ratio ($q \sim$ 0.2) and young age ($\tau \sim$
8\,Ma; it is a member of the \object{TW~Hydrae Association}) if compared to
those of the rest of exoplanetary systems, resembles more a recently-born,
low-mass, substellar binary than an exo-planetary system (Chauvin, priv.~comm.).
Another similar systems, some of them with wider projected physical separations
but without common proper motion confirmation, have also been discovered in the
\object{Orion OB1 association} (Caballero et~al. 2006; Barrado y Navascu\'es
et~al. 2007), \object{Upper Scorpius} (Kraus \& Hillenbrand 2007; B\'ejar et~al.
2008), and highly extinguished star-forming regions in the Southern Hemisphere,
such as Chamaeleon and Lupus (L\'opez Mart\'{\i} et~al. 2004, 2005; Luhman
2004). 

There remains five proper-motion-confirmed, low binding-energy
systems, of which three have total masses $M_1 + M_2 \ll$ 1\,$M_\odot$. 
They are the only representatives of the rare class of very wide
($s >$ 1000\,AU), very low-mass ($M_1 + M_2 \lesssim$ 0.2\,$M_\odot$),
equal-mass ($q \sim$ 1) binaries:
Koenigstuhl~1, 2M0126--50, and 2M1258+40.
The other two systems with total masses $M_1 + M_2 \sim$ 1--2\,$M_\odot$
have L-type companions at 3600 ($\eta$~CrB) and 11\,900\,AU (Koenigstuhl~3) to
bright stars.

Because of its relatively large value of $|U_g^*|$, typical of binaries in
the solar neighbourhood, the quintuple system HD~200077 in
Table~\ref{probableboundwidesystems} is likely bound. 
The same can be applied to HD~6101, $\alpha$~Cen, HD~136654, and $\xi^{02}$~Cap,
with values $|U_g^*| >$ 25\,10$^{33}$\,J.
However, both LP~268--35 and AU~Mic systems have very low absolute values
of potential energy, similar to those of the most weakly bound known
binaries, which are the Chauvin et~al. (2004) substellar pair and the three
binaries of very low-mass stars or brown dwarfs separated by more than
1000\,AU. 
LP~268--35 and AU~Mic, with projected physical separation between two and
three orders of magnitude larger, must be very fragile and will be soon
torn apart by third bodies, if they are not already in the process of
disruption.

\section{Summary}

Of the {104\,312} pairs in the Washington Double Star (WDS) catalogue (as in
2009 May), I~selected for follow-up the {36} pairs with tabulated angular
separations $\rho >$ 1000\,arcsec.
Of them, I~was not able to identify five, and a sixth pair had an actual angular
separation shorter than 1000\,arcsec.
I~rejected 12 of the remaining 30 pairs as binary candidated based on discordant
published proper motions, heliocentric distances, and radial velocities.
After a careful astro-photometric examination, with several astrometric epochs
covering at least 45\,a and proper-motion accuracies of 0.4--1.9\,mas\,a$^{-1}$,
only seven of the other 18 systems remained as probable bound systems. 
They were:
\begin{itemize}
\item WDS~01024+0504: a quadruple system containing HD~6101~AB, a K3V close
binary, and G~1--45~AB, a spectroscopic white dwarf binary.
\item WDS~13090+3353: LP~268--35 and LP~268--33, two poorly-know, late-type
dwarfs separated by about 84\,000\,AU. 
It is the widest (and most fragile) system in my sample.
The hypothetical secondary displays flux excess in the blue optical (SDSS $u$
and $g$) that can be ascribed to activity.   
\item WDS~14396--6050: the celebrated system $\alpha$~Cen~AB and Proxima.
\item WDS~15208+3129: a pair of F5V and K0V stars in the Hyades Supercluster.
I~first assign membership of the primary, HD~136654, in this moving group.
\item WDS~20124--1237: the bright star $\xi^{02}$~Cap and the M0Vk high-proper
motion star LP~754--50.
\item WDS~20452--3120: the very young stars AU~Mic and AT~Mic~AB in the
$\beta$~Pictoris moving group.
They have completed two orbital periods at most since their birth.
\item WDS~20599+4016: a hierarchical quintuple system around an F8V star.
With a total mass of about 4.1\,$M_\odot$, it is the most massive system in my
sample. 
\end{itemize}

Six of the seven hypothetical very wide systems have projected physical
separations larger than the cutoff at $s$ = 0.1\,pc stated in many classical
works, such as Bahcall \& Soneira (1981), Retterer \& King (1982), or Weinberg
et~al. (1987). 
Actually, there are four [two] systems with projected physical separations
larger than 0.2\,pc [0.3\,pc].
In order words, the cutoff at $s$ = 0.1\,pc is not an absolute limit: there are
systems with wider projected physical separations, although they are extremely
rare. 

Only two wide systems belong to young moving groups (Hyades Supercluster and
$\beta$~Pictoris), which indicates that the origin of such wide separations does
not only reside in the binary formation process, but also in the subsequent
dynamical evolution (e.g. by interaction with another stars in the
Galactic disc or with the interstellar medium).

All the systems except AU~Mic + AT~Mic~AB and LP~268--35 + LP~268--33
are consistent with being physical doubles.
To ascertain that, I~computed the minimum absolute values of binding energies,
$|U_g^*(s)|$ of the seven systems and compared them with those of a large
ensemble of systems containing at least one component less massive than
0.15\,$M_\odot$.  

There may exist bound systems wider than 10$^5$\,AU if they have enough
gravitational energy (i.e. total mass $M_1 + M_2 \gtrsim$ 6\,$M_\odot$), but
they will likely be young systems on the point of disruption by dynamical
encounters in the Galactic disc.

\begin{acknowledgements}

I thank the anonymous referee for his/her helpful report and careful
reading of the manuscript.
I~am an investigador Juan de la Cierva at the Universidad Complutense de Madrid.
This research has made use of the Washington Double Star catalog maintained at
the United States Naval Observatory, the SIMBAD, operated at Centre de Donn\'ees
astronomiques de Strasbourg, France, and the NASA's Astrophysics Data System.
Financial support was provided by the Universidad Complutense de Madrid,
the Comunidad Aut\'onoma de Madrid, the Spanish Ministerio Educaci\'on y
Ciencia, and the European Social Fund under grants:
AyA2005-02750,				
AyA2005-04286, AyA2005-24102-E,		
AyA2008-06423-C03-03, 			
AyA2008-00695,				
PRICIT S-0505/ESP-0237,			
and CSD2006-0070. 			

\end{acknowledgements}

\appendix

\section{Washington Double Stars with $\rho >$ 1000\,arcsec}

\begin{sidewaystable*}
\begin{minipage}[t][180mm]{\textwidth}
      \caption[]{Basic data from the literature of Washington Double Stars with 
      tabulated angular separations $\rho >$ 1000\,arcsec.} 
         \label{verywidebinaries.1}
\centering
         \begin{tabular}{lll cccc c ccc l cc}
            \hline
            \hline
            \noalign{\smallskip}
WDS		& Discovery  	& Simbad		& $\alpha$	& $\delta$      & $\mu_\alpha \cos{\delta}$	& $\mu_\delta$  	& $d_\pi$		& $V^{b}$ 	& $J$ 			& $K_{\rm s}$ 		& Sp.	 	& $\rho$	& $\theta$      \\ 
identifier  	& designation$^{a}$& name		& (J2000)	& (J2000)       & [mas\,a$^{-1}$]		& [mas\,a$^{-1}$]	& [pc]  		& [mag]		& [mag]			& [mag]			& type		& [arcsec]	& [deg]	        \\ 
            \noalign{\smallskip}
            \hline
            \noalign{\smallskip}
00059+1805	& LEP   1AE	& \object{HD 101}	& 00 05 54.74	& +18 14 05.8 	& --152.2$\pm$1.1		& --148.1$\pm$1.4	& 37.1$\pm$0.8	 	& 7.46 		& 6.332$\pm$0.019	& 6.034$\pm$0.024	& F8		& 1606.8	& 245.9	        \\ 
		&		& \object{LP 404--21}	& 00 04 11.81	& +18 03 10.7	& --146				& --146			& ...		  	& 16.3		& 14.068$\pm$0.030	& 13.339$\pm$0.029	& ...		&		& 	        \\ 
            \noalign{\smallskip}
00152+2454	& GIC   4	& \object{G 131--46}	& 00 16 53.59	& +24 20 48.7  	& --136$\pm$11			& --204$\pm$10		& ...		 	& 13.6		& 9.944$\pm$0.020 	& 9.067$\pm$0.017 	& K		& 2469.8	& 145.8         \\ 
		&		& \object{G 130--59}	& 00 15 11.85	& +24 54 52.1	& --105$\pm$11			& --215$\pm$11		& ...		  	& 14.5		& 12.600$\pm$0.022 	& 11.951$\pm$0.022 	& ...		&		&	        \\ 
           \noalign{\smallskip}
00400--1533	& LDS	5286	& \object{LP 765--52}	& 00 37 29.44	& --15 45 52.3  & +173$\pm$5	       		& --19$\pm$5       	& ...		 	& 15.5 *	& 11.890$\pm$0.021 	& 11.038$\pm$0.023 	& ...		& 2274.9	& 71.5	        \\ 
		&		& \object{LP 765--57}	& 00 39 59.21	& --15 34 07.9	& +169$\pm$5	       		& --34$\pm$5       	& ...		  	& 15.2 *	& 13.317$\pm$0.026 	& 12.746$\pm$0.035 	& ...		&		&	        \\ 
           \noalign{\smallskip}
00435+3351	& WEI  46AC	& \object{GJ 30}	& 00 43 32.90	& +33 50 41.3	& --204.0$\pm$0.8	       	& --357.2$\pm$0.6       & 20.6$\pm$0.4	 	& 8.73		& 6.592$\pm$0.018 	& 5.938$\pm$0.016 	& K8		& 1295.0	& 266.8	        \\ 
		&		& \object{BD+33 96}	& 00 41 49.11	& +33 51 54.8	& +8.3$\pm$1.2	       		& +1.8$\pm$1.2       	& ...		  	& 8.63		& 6.639$\pm$0.021 	& 5.966$\pm$0.018 	& K0		&		&	        \\ 
           \noalign{\smallskip}
00520+2035	& GIC	16	& \object{G 69--27}	& 00 52 00.03	& +20 34 58.6	& +176.5$\pm$1.8	       	& --107.4$\pm$1.5       & 32$\pm$3	 	& 11.40		& 8.490$\pm$0.019 	& 7.628$\pm$0.020 	& ...		& 1614.0	& 56.1	        \\ 
		&		& \object{G 69--29}	& 00 53 35.48	& +20 49 59.4	&   +213$\pm$11		       	&   --132$\pm$11       	& ...		  	& 13.9 *	& 10.095$\pm$0.021 	& 9.248$\pm$0.015 	& ...		&		&	        \\ 
           \noalign{\smallskip}
01024+0504	& WNO  50AC	& \object{HD 6101} AB	& 01 02 24.60	& +05 03 41.4	& +323.3$\pm$1.2	       	& +226.0$\pm$1.2       	& 21.1$\pm$0.5	 	& 8.16		& 6.199$\pm$0.019 	& 5.510$\pm$0.020 	& K2+...	& 1276.0	& 87.8	        \\ 
		&		& \object{G~1--45} AB	& 01 03 49.94	& +05 04 30.7	&   +317$\pm$8		       	&   +227$\pm$8       	& 21.3$\pm$1.7		& 14.10		& 13.504$\pm$0.024 	& 13.418$\pm$0.034 	& DA5		&		&	        \\ 
           \noalign{\smallskip}
01163--3217	& LDS	1091	& \object{LP 883--336}	& 01 16 21.75	& --31 58 12.2	& +408$\pm$5	       		& --1$\pm$5	       	& ...		 	& 17.9 *	& 11.899$\pm$0.022 	& 11.001$\pm$0.023 	& ...		& 1151.0	& 186.0	        \\ 
		&		& \object{LP 883--337}	& 01 16 12.35	& --32 17 17.0	& +388$\pm$5	       		& +17$\pm$5	       	& ...		  	& 18.0 *	& 13.884$\pm$0.026 	& 13.132$\pm$0.034 	& ...		&		&	        \\ 
           \noalign{\smallskip}
02255--0904	& GRV	1148	& \object{WD 0223--092}	& 02 25 30.96	& --09 04 14.9	& +81$\pm$5	       		& --2$\pm$5	       	& ...		 	& 19.5 *	& ...		 	& ...		 	& DA4.6		& 1272.1	& 228.2	        \\ 
		&		& \object{WD 0221--095}	& 02 24 26.96	& --09 18 23.5	& +85$\pm$4	       		& --0$\pm$4	       	& ...		  	& 19.8 *	& ...		 	& ...		 	& DA5.8		&		&	        \\ 
           \noalign{\smallskip}
02310+0823	& GIC	32	& \object{G 73--63}	& 02 31 03.28	& +08 22 55.1	& +376.1$\pm$1.9	       	& --85.4$\pm$1.6	& 33$\pm$3	 	& 10.90		& 8.356$\pm$0.023 	& 7.554$\pm$0.023 	& K4--7V:	& 3094.6	& 277.9	        \\ 
		&		& \object{G 73--59}	& 02 27 36.68	& +08 29 58.8	& +350		       		& --63		       	& ...		  	& 16.1 *	& 11.271$\pm$0.026 	& 10.455$\pm$0.021 	& ...		&		&	        \\ 
           \noalign{\smallskip}
03162+5810	& LEP	13AC	& \object{GJ 130.1 A}	& 03 16 13.82	& +58 10 02.4 	& +445.6$\pm$3.9	       	& --340.3$\pm$4.1	& 14.4$\pm$0.7	 	& 10.53		& 7.344$\pm$0.020 	& 6.566$\pm$0.024 	& M2 		& 1164.1	& 197.6	        \\ 
		&		& \object{G 246--30}	& 03 15 29.44 	& +57 51 33.0	& +467		       		& --237		       	& ...		  	& 15.5 *	& 11.121$\pm$0.024 	& 10.271$\pm$0.019 	& M:		&		&	        \\ 
           \noalign{\smallskip}
03330+0306	& LDS 	3504	& \object{G 80--8}	& 03 32 59.02	& +03 06 08.0 	& +294.8			& --80.9		& ...		  	& 13.1 *	& 10.102$\pm$0.023 	& 9.281$\pm$0.023 	& ... 		& 429.5		& 266.4	        \\ 
		&		& \object{NLTT 11184}	& 03 32 30.40 	& +03 05 40.8	& +286		       		& --77		       	& ...		  	& 17.0 *	& 12.491$\pm$0.026 	& 11.750$\pm$0.029 	& ...		&		&	        \\ 
           \noalign{\smallskip}
03442--6448	& LDS 	104	& \object{$\beta$ Ret} AB& 03 44 11.96	& --64 48 24.9 	& +310.1$\pm$0.7	       	& +83.2$\pm$0.6		& 29.9$\pm$0.5	 	& 3.84		& 1.937$\pm$0.310 	& 1.279$\pm$0.270 	& K2III SB 	& 1466.3	& 94.2	        \\ 
		&		& \object{HD 24293}	& 03 48 01.12 	& --64 50 11.7	& +334.0$\pm$1.0	       	& +99.0$\pm$1.1		& 40.5$\pm$1.3	 	& 7.85		& 6.630$\pm$0.029 	& 6.241$\pm$0.024 	& G3V		&		&	        \\ 
           \noalign{\smallskip}
07590--6338	& LDS 199	& \object{CD--63 370}	& 07 58 57.36	& --63 37 45.5	& --143.0$\pm$1.7	       	& +262.5$\pm$1.8        & ...	 		& 9.90		& 8.826$\pm$0.034 	& 8.502$\pm$0.023 	& F8		& 1033.4	& 195.8		\\ %
		&		& \object{L 137--85}	& 07 58 14.99	& --63 54 19.6	& --178$\pm$10		       	& +384$\pm$10       	& ...		  	& 11.2 *	& 9.791$\pm$0.024 	& 9.175$\pm$0.023 	& ...		&		&	        \\ %
           \noalign{\smallskip}
10197+1928	& WNO  53	& \object{40~Leo}	& 10 19 44.20	& +19 28 15.8	& --230.0$\pm$0.9	       	& --214.7$\pm$0.5       & 21.37$\pm$0.11	& 4.78		& 4.037$\pm$0.292 	& 4.020$\pm$0.314 	& F6IV		& 5231.4	& 308.3		\\ 
		&		& \object{LP~371--59} A & 10 14 53.94	& +20 22 18.9	& --225			       	& --198       		& ...		  	& 15.3 *	& 10.815$\pm$0.026 	& 9.99$\pm$0.023 	& M5		&		&	        \\ %
           \noalign{\smallskip}
11125+3549	& STTA	108BD	& \object{HD 97371}	& 11 12 44.28	& +35 49 48.4 	& --60.1$\pm$0.7	       	& --7.8$\pm$0.7		& 141$\pm$13	 	& 7.20		& 5.423$\pm$0.019 	& 4.805$\pm$0.018 	& K0	 	& 2100.8	& 88.5	        \\ 
		&		& \object{HD 97832}	& 11 15 36.98 	& +35 50 44.7	&   +0.1$\pm$1.2		& --2.2$\pm$1.3		& ...		 	& 8.20		& 6.668$\pm$0.021 	& 6.194$\pm$0.018 	& G5		&		&	        \\ 
           \noalign{\smallskip}
11452+1821	& GIC	101	& \object{G 57--17}	& 11 45 11.92	& +18 20 58.7	& --296$\pm$8			& --296$\pm$8		& ...		 	& 13.27		& 9.162$\pm$0.022 	& 8.260$\pm$0.016 	& M4		& 1341.3	& 280.2	        \\ 
		&		& \object{G 57--15}	& 11 43 39.18	& +18 24 56.9	& --259		       		& --267		       	& ...		  	& 15.07		& 12.863$\pm$0.026 	& 12.097$\pm$0.021 	& ...		&		&	        \\ 
           \noalign{\smallskip}
            \hline
         \end{tabular}
\vfill
\end{minipage}
\end{sidewaystable*}

\begin{sidewaystable*}
\begin{minipage}[t][180mm]{\textwidth}
      \caption[]{Basic data from the literature of Washington Double Stars with 
      tabulated angular separations $\rho >$ 1000\,arcsec.} 
         \label{verywidebinaries.2}
\centering
         \begin{tabular}{lll cccc c ccc l cc}
            \hline
            \hline
            \noalign{\smallskip}
WDS		& Discovery  	& Simbad		& $\alpha$	& $\delta$      & $\mu_\alpha \cos{\delta}$	& $\mu_\delta$  	& $d_\pi$		& $V^{b}$ 	& $J$ 			& $K_{\rm s}$ 		& Sp.	 	& $\rho$	& $\theta$      \\ 
identifier  	& name$^{a}$	& name			& (J2000)	& (J2000)       & [mas\,a$^{-1}$]		& [mas\,a$^{-1}$]	& [pc]  		& [mag]		& [mag]			& [mag]			& type		& [arcsec]	& [deg]	        \\ 
            \noalign{\smallskip}
            \hline
            \noalign{\smallskip}
11455+4740	& LEP	45	& \object{HD 102158}	& 11 45 30.58	& +47 40 01.1	& --591.6$\pm$0.7	       	& --290.7$\pm$0.5	& 49.3$\pm$1.7	 	& 8.06		& 6.860$\pm$0.026 	& 6.509$\pm$0.026 	& G2V		& 1176.1	& 72.4	        \\ 
		&		& \object{G 122--46}	& 11 47 21.66 	& +47 45 56.7	& --585		       		& --200		       	& ...		  	& 14.16		& 10.586$\pm$0.020 	& 9.846$\pm$0.020 	& M:		&		&	        \\ 
           \noalign{\smallskip}
13090+3353	& LEP  62AC	& \object{LP 268--35}	& 13 08 58.26	& +33 53 10.0	& --209.7$\pm$2.3	       	& --111.8$\pm$1.7	& 66$\pm$12	 	& 11.56		& 9.296$\pm$0.019 	& 8.623$\pm$0.018 	& ...		& 1273.8	& 192.3	        \\ 
		&		& \object{LP 268--33}	& 13 08 36.42 	& +33 32 25.7	& --208.8$\pm$5.5		& --101.2$\pm$5.5	& ...		  	& 16.3 *	& 11.685$\pm$0.021 	& 10.769$\pm$0.019 	& ...		&		&	        \\ 
           \noalign{\smallskip}
13410+6808	& LDS	5788	& \object{G 238--50}	& 13 44 29.36 	& +68 27 50.3	& --269.9$\pm$1.5	       	& +43.5$\pm$1.8		& 40$\pm$3	 	& 11.17		& 8.845$\pm$0.021 	& 8.023$\pm$0.015 	& M2		& 1696.8	& 44.5	        \\ 
		&		& \object{LP 40--200}	& 13 40 54.27	& +68 07 43.8	& --265.5$\pm$1.8		& +32.8$\pm$1.8		& ...		  	& 11.83		& 9.837$\pm$0.019 	& 9.138$\pm$0.020 	& K3		&		&	        \\ 
           \noalign{\smallskip}
13599+2520	& BUP	156	& \object{BD+26 2517} AB& 14 01 11.60   & +25 21 36.2   & +5.5$\pm$1.4			& +36.2$\pm$1.3	  	& ... 		  	& 9.51	  	& 8.337$\pm$0.023	& 8.060$\pm$0.018	& G0  	  	& 1033.9	& 83.2	        \\ 
		&		& \object{BD+26 2513}	& 13 59 55.87   & +25 19 33.7   & +20.5$\pm$1.2			& --55.9$\pm$1.2	& ... 		  	& 10.41	  	& 9.433$\pm$0.021	& 9.145$\pm$0.019	& F8  	  	&		&	        \\ 
           \noalign{\smallskip}
14396--6050$^{c}$& LDS	494AC	& \object{$\alpha$ Cen} AB& 14 39 35.93 & --60 50 07.0 	& --3633.7$\pm$0.7		& +702.3$\pm$0.6  	& 1.325$\pm$0.007	& 1.35		& --1.454$\pm$0.133	& --2.008$\pm$0.260	& G2V+K1V 	& 7860.1	& 236.9	        \\ 
		&		& \object{Proxima}	& 14 29 42.91   & --62 40 46.5  & --3775.8$\pm$1.6		& +766$\pm$2	  	& 1.296$\pm$0.004	& 11.01	  	& 5.357$\pm$0.023	& 4.384$\pm$0.033	& M5.5Ve  	&		&	        \\ 
           \noalign{\smallskip}
15208+3129	& LEP	74	& \object{HD 136654}	& 15 20 50.07  	& +31 28 48.5  	& --180.0$\pm$0.4		& +139.0$\pm$0.5	& 43.5$\pm$1.1		& 6.90	  	& 5.982$\pm$0.020	& 5.742$\pm$0.027	& F5V	    	& 1580.6	& 325.6	        \\ 
		&		& \object{BD+32 2572}	& 15 19 40.15   & +31 50 32.9  	& --181.2$\pm$0.6		& +140.2$\pm$0.8	& 42$\pm$2		& 9.03	  	& 7.569$\pm$0.021	& 7.115$\pm$0.020	& K0V	  	&		&	        \\ 
           \noalign{\smallskip}
16348--0412	& GIC	144AB	& \object{HD 149414} AB	& 16 34 42.36  	& --04 13 44.1  & --133.7$\pm$1.4		& --701.2$\pm$1.4	& 45$\pm$3		& 9.60	  	& 8.055$\pm$0.024	& 7.517$\pm$0.024	& G5Ve SB$_1$	& 1176.5	& 36.4	        \\ 
		&		& \object{BD--03 3968B}	& 16 35 29.03 	& --03 57 57.2  & --162				& --685			& ... 		  	& 13.86	  	& 11.086$\pm$0.022	& 10.541$\pm$0.022	& M:	  	&		&	        \\ 
           \noalign{\smallskip}
18111+3241	& LEP	87	& \object{BD+32 3065}	& 18 11 06.22  	& +32 41 00.3	& --134.7$\pm$1.2		& +322.1$\pm$1.3	& 46$\pm$3		& 10.52	  	& 8.590$\pm$0.039	& 7.931$\pm$0.020	& K5		& 1106.0	& 148.3	        \\ 
		&		& \object{G 206--16}	& 18 11 52.28 	& +32 25 20.0  	& --143				& +318			& ... 		  	& 15.6 *	& 10.885$\pm$0.021	& 10.024$\pm$0.017	& ...	  	&		&	        \\ 
           \noalign{\smallskip}
20084+1503	& LDS	1033AF	& \object{G 143--33}	& 20 08 21.93  	& +15 02 36.6	& --159.1$\pm$1.7		& --180.7$\pm$1.7	& ...			& 11.56	  	& 9.408$\pm$0.028	& 8.409$\pm$0.024	& ...		& 2186.6	& 269.0	        \\ 
		&		& \object{G 143--27}	& 20 05 51.01 	& +15 01 59.2  	& --156				& --210			& ... 		  	& 12.88		& 11.626$\pm$0.022	& 11.245$\pm$0.020	& ...	  	&		&	        \\ 
           \noalign{\smallskip}
20124--1237	& TDT	2085AC	& \object{$\xi^{02}$ Cap}& 20 12 25.86  	& --12 37 02.5	& +193.4$\pm$0.4		& --196.0$\pm$0.5	& 27.7$\pm$0.3		& 5.84	  	& 4.971$\pm$0.020	& 4.634$\pm$0.017	& F7V		& 1021.3	& 193.6	        \\ 
		&		& \object{LP 754--50}	& 20 12 09.44 	& --12 53 35.1  & +195.3$\pm$2.0		& --194.8$\pm$1.8	& 23.6$\pm$1.4		& 11.30		& 8.485$\pm$0.023	& 7.625$\pm$0.021	& M0Vk	  	&		&	        \\ 
           \noalign{\smallskip}
20302+2651	& BUP	213AE	& \object{HD 340345} AB	& 20 30 10.67  	& +26 50 34.5	& --145.9$\pm$2.7		& --141.9$\pm$2.7	& 23.8$\pm$1.8		& 9.69	  	& 7.133$\pm$0.021	& 6.347$\pm$0.020	& M1V+		& 1338.7	& 93.2	        \\ 
		&		& \object{HD 340459}	& 20 31 50.53 	& +26 49 19.3  	& --8.1$\pm$1.4			& --5.5$\pm$1.4		& ... 		  	& 10.15		& 8.235$\pm$0.023	& 7.647$\pm$0.016	& G5	  	&		&	        \\ 
           \noalign{\smallskip}
20452--3120	& LDS	720AB	& \object{AU Mic}	& 20 45 09.49  	& --31 20 26.7	& +279.6$\pm$1.2		& --360.3$\pm$0.8	& 9.92$\pm$0.10		& 8.81	  	& 5.436$\pm$0.017	& 4.529$\pm$0.020	& M1Ve		& 4680.9	& 212.8	        \\ 
		&		& \object{AT Mic} AB	& 20 41 51.12 	& --32 26 07.3  & +261.3$\pm$3.6		& --344.8$\pm$3.9	& 10.7$\pm$0.4		& 10.27		& 5.807$\pm$0.026	& 4.944$\pm$0.042	& M4Ve+M5:	&		&	        \\ 
           \noalign{\smallskip}
20599+4016	& LEP  98AD	& \object{HD 200077} AE--D& 20 59 55.24	& +40 15 31.4	& +230.5$\pm$1.1		& +211.2$\pm$1.3	& 41.0$\pm$0.9		& 6.58	  	& 5.450$\pm$0.021	& 5.119$\pm$0.024	& F8+		& 1212.8	& 258.5	        \\ 
		&		& \object{G 210--44} AB	& 20 58 11.48 	& +40 11 29.0   & +229.7$\pm$1.2		& +202.9$\pm$1.2	& 46$\pm$4		& 10.75		& 8.142$\pm$0.030	& 7.339$\pm$0.018	& M1		&		&	        \\ 
           \noalign{\smallskip}
22175+2335	& GIC 	179	& \object{G 127--13}	& 22 17 25.87  	& +23 35 04.7	& --104$\pm$12			& --388$\pm$11		& ... 		  	& 14.1 *	& 9.890$\pm$0.019	& 9.057$\pm$0.023	& ...		& 2107.7	& 12.7	        \\ 
		&		& \object{G 127--14}	& 22 17 59.56 	& +24 09 21.1   & --124$\pm$10			& --416$\pm$10		& ... 		  	& 14.2 *	& 10.276$\pm$0.021	& 9.498$\pm$0.018	& ...		&		&	        \\ 
           \noalign{\smallskip}
23228+2208	& GIC 	191	& \object{BD+21 4923} 	& 23 22 48.81  	& +22 07 59.4	& +198.3$\pm$1.2		& --69.8$\pm$1.2	& 71$\pm$9		& 9.71		& 8.293$\pm$0.020	& 7.851$\pm$0.016	& F5		& 3265.3	& 253.9	        \\ 
		&		& \object{G 68--7}	& 23 19 03.21 	& +21 52 54.5   &   +310$\pm$11			&  --101$\pm$11		& ... 		  	& 14.7 *	& 11.589$\pm$0.020	& 10.784$\pm$0.018	& ...		&		&	        \\ 
           \noalign{\smallskip}
            \hline
         \end{tabular}
\begin{list}{}{}
\item[$^{a}$] References of discovery names -- 
BUP: Burnham, S.~W. proper motion stars (from additional ``DD'' list):
GIC: Giclas et~al. (1961); 
GRV: Greaves, J. private communication (from data in Eisenstein et~al. 2006);
LEP: L\'epine \& Bongiorno (2007); 
LDS: Luyten, W.~J. proper motion catalogues (e.g. Luyten 1941); 
STF: Struve, F.~J.~W. (several citations by e.g.: Herschel 1833; Bessel 1833);
STTA: Struve, O. ``DD'' (Appendix list);
TDT: Tycho Double Star (from additional ``DD'' list);
WNO: Washington Observations;
WEI: Weisse, M. (cited in, e.g., Burnham Double Star Catalogue 1906); 
\item[$^{b}$] $V$-band magnitudes marked with an asterisk are estimated from 
four epochs of photographic $B_J$ and $R_F$ magnitudes or from Sloan $g$ and $r$
magnitudes.
Remaining $V$-band magnitudes are from the literature (see main text).
\item[$^{c}$] The proper motion of the primary in the system corresponds to
$\alpha$~Cen~A. 
The apparent difference with respect Proxima (of up to 292$\pm$4\,mas\,a$^{-1}$
in $\mu_\delta$) is due to the strong dynamical effect on $\alpha$~Cen~A by its
close companion, $\alpha$~Cen~B. 
\end{list}
\vfill
\end{minipage}
\end{sidewaystable*}


\begin{thebibliography}{}

\bibitem[1988]{Abt88} Abt, H. A. 1988, ApJ, 331, 922

\bibitem[2008]{AdC08} Adelman-McCarthy, J., Ag\"ueros, M. A., Allam, S. S. 2008,
ApJS, 175, 297

\bibitem[2000]{APH00} Allen, C., Poveda, A., Herrera, M. A. 2000, A\&A, 356, 529

\bibitem[1991]{AO91} Anosova, J. P., Orlov, V. V. 1991, A\&A, 252, 123

\bibitem[1994]{AOP94} Anosova, J. P., Orlov, V. V., Pavlova, N. A. 1994, A\&A,
292, 115

\bibitem[2007]{Ar07} Artigau, \'E., Lafreni\`ere, D., Doyon, R. et~al. 2007,
ApJ, 659, L49

\bibitem[2009]{Ar09} Artigau, \'E., Lafreni\`ere, D., Albert, L., Doyon, R.
2009, ApJ, 692, 149

\bibitem[1981]{BS81} Bahcall, J. N., Soneira, R. M. 1981, ApJ, 246, 122

\bibitem[2002]{Bak02} Bakos, G. \'A., Sahu, K. C., N\'emeth, P. 2002, ApJS, 141,
187

\bibitem[2002]{Bal02} Balega, I. I., Balega, Y. Y., Hofmann, K.-H. et~al. 2002,
A\&A, 385, 87

\bibitem[2004]{Bal04} Balega, I. I., Balega, Y. Y., Maksimov, A. F. et~al. 2004,
A\&A, 422, 627

\bibitem[2006]{Bal06} Balega, I. I., Balega, Y. Y., Hofmann, K.-H. et~al. 2006,
A\&A, 448, 703

\bibitem[2007]{Bal07} Balega, I. I., Balega, Y. Y., Maksimov, A. F. et~al. 2007,
AstBu, 62, 339

\bibitem[1998]{Bar98} Baraffe, I., Chabrier, G., Allard, F., Hauschildt, P. H.
1998, A\&A, 337, 403

\bibitem[1999]{ByN99} Barrado y Navascu\'es, D., Stauffer, J. R., Song, I.,
Caillault, J.-P. 1999, ApJ, 520, L123

\bibitem[2007]{ByN07} Barrado y Navascu\'es, D., Bayo, A., Morales-Calder\'on,
M. et~al. 2007, A\&A, 468, L5

\bibitem[1996]{Bat96} Batalha, C. C., Stout-Batalha, N. M., Basri, G., Terra, M.
A. O. 1996, ApJS, 103, 211

\bibitem[1973]{Bat73} Batten, A.~H. 1973, Binary and multiple systems of stars,
Oxford, New York, Pergamon Press (International series of monographs in natural
philosophy; vol.~51) 

\bibitem[2006]{Be06} Beaulieu, J.-P., Bennett, D. P., Fouqu\'e, P. et~al. 2006,
Nature, 439, 437

\bibitem[2008]{Be08} B\'ejar, V. J. S., Zapatero Osorio, M. R., P\'erez-Garrido,
A. et~al. 2008, ApJ, 673, L185

\bibitem[2001]{Be01} Bergeron, P., Leggett, S. K., Ruiz, M. T. 2001, ApJS, 133,
413

\bibitem[1833]{Be1833} Bessel, F. W. 1833, AN, 10, 389

\bibitem[2007]{Bo07} Bochanski, J. J., West, A. A., Hawley, S. L., Covey, K. R.
2007, AJ, 133, 531

\bibitem[2000]{Bo00} Bonnarel, F., Fernique, P., Bienaym\'e, O. et~al. 2000,
A\&AS, 143, 3

\bibitem[2003]{Bo03} Bouy, H., Brandner, W., Mart\'{\i}n, E. L. et~al. 2003, AJ,
126, 1526

\bibitem[2005]{Bo05} Bouy, H., Mart\'{\i}n, E. L., Brandner, W., Bouvier, J.
2005, AN, 326, 969

\bibitem[2006]{BM06} Burgasser, A. J., McElwain, M. W. 2006, AJ, 131, 1007

\bibitem[2000]{Bu00} Burgasser, A. J., Kirkpatrick, J. D., Cutri, R. M. et~al.
2000, ApJ, 531, L57

\bibitem[2007]{Bu07} Burgasser, A. J., Reid, I. N., Siegler, N. et~al. 2007,
Protostars and Planets V, eds. B. Reipurth, D. Jewitt, and K. Keil, University
of Arizona Press, Tucson, 2007, p.~427 

\bibitem[1906]{Bu1906} Burnham, S. W. 1906, A general catalogue of double stars
within 121\,deg of the North pole, Carnegie institution of Washington,
University of Chicago press 

\bibitem[2007]{Ca07a} Caballero, J.~A. 2007a, A\&A, 466, 917

\bibitem[2007]{Ca07b} Caballero, J.~A. 2007b, ApJ, 667, 520

\bibitem[2008]{CD08} Caballero, J.~A., Dinis, L. 2008, AN, 329, 801

\bibitem[2006]{Ca06} Caballero, J.~A., Mart\'{\i}n, E. L., Dobbie, P. D.,
Barrado y Navascu\'es, D. 2006, A\&A, 460, 635

\bibitem[2008]{Ca08} Caballero, J.~A., Burgasser, A. J., Klement, R. 2008, A\&A,
488, 181

\bibitem[1994]{Ca04} Carney, B. W., Latham, D. W., Laird, J. B., Aguilar, L. A.
1994, AJ, 107, 2240

\bibitem[2004]{CG04} Chanam\'e, J., Gould, A. 2004, ApJ, 601, 289

\bibitem[2004]{Ch04} Chauvin, G., Lagrange, A.-M., Dumas, C. et~al. 2004, A\&A,
425, L29

\bibitem[2005]{Ch05} Chauvin, G., Lagrange, A.-M., Dumas, C. et~al. 2005, A\&A,
438, L25

\bibitem[2001]{Ch01} Chen, Y. Q., Nissen, P. E., Benoni, T., Zhao, G. 2001,
A\&A, 371, 943

\bibitem[1990]{Cl90} Close, L. M., Richer, H. B., Crabtree, D. R. 1990, AJ, 100,
1968

\bibitem[2002]{Cu02} Cutispoto, G., Pastori, L., Pasquini, L. et~al. 2002, A\&A,
384, 491

\bibitem[2006]{Ei06} Eisenstein, D. J., Liebert, J., Harris, H. C. et~al. 2006,
ApJS, 167, 40

\bibitem[1967]{Ev67} Evans, D. S. 1967, 
Determination of Radial Velocities and their Applications. 
Proceedings from International Astronomical Union Symposium no. 30.
University of Toronto. 
20--24 June 1966. 
Ed. A.~H. Batten \& J.~F. Heard. 
Academic Press, London, p.~57

\bibitem[2002]{Ev02} Evans, D. W., Irwin, M. J., Helmer, L. 2002, A\&A,
395, 347

\bibitem[2005]{FBZ05} Farihi, J., Becklin, E. E., Zuckerman, B. 2005, ApJS, 161,
394

\bibitem[2005]{FV05} Fischer, D. A., Valenti, J. 2005, ApJ, 622, 1102

\bibitem[2005]{Fo05} Forveille, T., Beuzit, J.-L., Delorme, P. et~al. 2005,
A\&A, 435, L5

\bibitem[1966]{Ga66} Gasteyer, C. 1966, AJ, 71, 1017

\bibitem[1959]{Gi59} Giclas, H. L., Slaughter, C. D., Burnham, R. 1959, LowOB,
4, 136

\bibitem[1961]{Gi61} Giclas, H. L., Burnham, R., Thomas, N. R. 1961, LowOB, 5,
61

\bibitem[2000]{Gi00} Gizis, J. E., Monet, D. G., Reid, I. N. et~al. 2000, AJ,
120, 1085 

\bibitem[2001]{Gi01} Gizis, J. E., Kirkpatrick, J. D., Burgasser, A. et~al.
2001, ApJ, 551, L163

\bibitem[2003]{Gi03} Gizis, J. E., Reid, I. N., Knapp, G. R. et~al. 2003, AJ,
125, 3302

\bibitem[1969]{Gl69} Gliese, W. 1969, Ver\"offentlichungen des Astronomischen
Rechen-Instituts Heidelberg, Nr. 22, ed. G. Braun, Karlsruhe

\bibitem[1988]{GJ88} Gliese, W., Jahreiss, H. 1988, Ap\&SS, 142, 49

\bibitem[2002]{Go02} Goldberg, D., Mazeh, T., Latham, D. W. et~al. 2002, AJ,
124, 1132

\bibitem[1999]{Go99} Goldman, B., Delfosse, X., Forveille, T. et~al. 1999, A\&A,
351, L5

\bibitem[2003]{Gr03} Gray, R. O., Corbally, C. J., Garrison, R. F. et~al. 2003,
AJ, 126, 2048 

\bibitem[2006]{Gr06} Gray, R. O., Corbally, C. J., Garrison, R. F. et~al. 2006,
AJ, 132, 161

\bibitem[1986]{Gr86} Green, R. F., Schmidt, M., Liebert, J. 1986, ApJS, 61, 305

\bibitem[2001]{Ha01} Hambly, N. C., MacGillivray, H. T., Read, M. A. et~al.
2001, MNRAS, 326, 1279

\bibitem[1993]{Ha93} Harrington, R. S., Dahn, C. C., Kallarakal, V. V. et~al.
1993, AJ, 105, 1571

\bibitem[1783]{He1783} Herschel, W. 1783, RSPT, 73, 4

\bibitem[1833]{He1833} Herschel, J. F. W. 1833, RSPT, 123, 359

\bibitem[2000]{Hog00} H{\o}g, E., Fabricius, C., Makarov, V. V. et~al. 2000,
A\&A, 355, L27

\bibitem[1915]{In15} Innes, R. T. A. 1915, Union Obs, Circ., 30

\bibitem[1974]{JA84} Joy, A.~H. \& Abt H.~A. 1974, ApJS, 28, 1

\bibitem[2002]{Kai02} Kaiser, N., Aussel, H., Burke, B. E. et~al. 2002, SPIE,
4836, 154

\bibitem[2004]{Kal04} Kalas, P., Liu, M. C., Matthews, B. C. 2004, Science, 303,
1990 

\bibitem[1978]{KW78} Kamper, K. W., Wesselink, A. J. 1978, AJ, 83, 1653

\bibitem[2004]{Kar04} Karata{\c s}, Y., Bilir, S., Eker, Z., Demircan, O. 2004,
MNRAS, 349, 1069

\bibitem[1999]{Kat99} Katsova, M. M., Drake, J. J., Livshits, M. A. 1999, ApJ,
510, 986

\bibitem[2001]{Ki01} Kirkpatrick, J. D., Dahn, C. C., Monet, D. G. et~al. 2001,
AJ, 121, 3235

\bibitem[1985]{Kr85} Kraicheva, Z. T., Popova, E. I., Tutukov, A. V., Iungelson,
L. R. 1985, Afz, 22, 105

\bibitem[2007]{Kr07} Kraus, A. L., Hillenbrand, L. A. 2007, ApJ, 662, 413

\bibitem[1987]{Ku87} Kundu, M. R., Jackson, P. D., White, S. M., Melozzi, M.
1987, ApJ, 312, 822

\bibitem[2009]{La09} Lagrange, A.-M., Desort, M., Galland, F., Udry, S., Mayor,
M. 2009, A\&A, 495, 335

\bibitem[2007]{LaB07} Lajoie, C.-P., Bergeron, P. 2007, ApJ, 667, 1126

\bibitem[2004]{LR04} Lambert, D. L., Reddy, B. E. 2004, MNRAS, 349, 757

\bibitem[2001]{La01} Lane, B. F., Zapatero Osorio, M. R., Britton, M. C., et~al.
2001, ApJ, 560, 390 

\bibitem[1988]{La88} Latham, D. W., Mazeh, T., Carney, B. W. et~al. 1988, AJ,
96, 567 

\bibitem[1991]{La91} Latham, D. W., Davis, R. J., Stefanik, R. P. et~al. 1991,
AJ, 101, 625 

\bibitem[2002]{La02} Latham, D. W., Stefanik, R. P., Torres, G. et~al. 2002, AJ,
124, 1144

\bibitem[2005]{LS05} L\'epine, S., Shara, M. M. 2005, AJ, 129, 1483

\bibitem[2007]{LeB07} L\'epine, S., Bongiorno, B. 2007, AJ; 133, 889

\bibitem[1847]{LV1847} Le~Verrier, U. J. 1847, AN, 25, 85

\bibitem[1988]{LDM88} Liebert, J., Dahn, C. C., Monet, D. G. 1988, ApJ, 332, 891

\bibitem[1982]{Li82} Linsky, J. L., Bornmann, P. L., Carpenter, K. G. et~al.
1982, ApJ, 260, 670

\bibitem[2004]{Lo04} L\'opez Mart\'{\i}, B., Eisl\"offel, J., Scholz, A., Mundt,
R. 2004, A\&A, 416, 555

\bibitem[2005]{Lo05} L\'opez Mart\'{\i}, B., Eisl\"offel, J., Mundt, R. 2005,
A\&A, 440, 139 

\bibitem[2004]{Lu04} Luhman, K. L. 2004, ApJ, 614, 318

\bibitem[1941]{Lu41} Luyten, W. J. 1941, Bruce proper motion survey, Minneapolis

\bibitem[2008]{Mak08} Makarov, V. V., Zacharias, N., Hennessy, G. S. 2008, ApJ,
687, 566

\bibitem[1999]{Mas99} Mason, B. D., Martin, C., Hartkopf, W. I. et~al. 1999, AJ,
117, 1890

\bibitem[2001]{Mas01} Mason, B. D., Wycoff, G, L., Hartkopf, W. I. et~al. 2001,
AJ, 122, 3466 

\bibitem[1993]{MG93} Matthews, R., Gilmore, G. 1993, MNRAS, 261, L5

\bibitem[2000]{Max00} Maxted, P. F. L., Marsh, T. R., Moran, C. K. J. 2000,
MNRAS, 319, 305

\bibitem[2003]{Maz03} Mazeh, T., Simon, M., Prato, L. et~al.
2003, ApJ, 599, 1344

\bibitem[1987]{Mc87} McAlister, H. A., Hartkopf, W. I., Hutter, D. J. et~al.
1987, AJ, 93, 183 

\bibitem[1999]{MS99} McCook, G. P., Sion, E. M. 1999, ApJS, 121, 1

\bibitem[1976]{Mc76} McMillan, R. S., Breger, M., Ferland, G. J., Loumos, G. L.
1976, PASP, 88, 495

\bibitem[2003]{Mo03} Monet, D. G., Levine, S. E., Canzian, B. et~al. 2003, AJ,
125, 984

\bibitem[2001]{Mo01} Montes, D., L\'opez-Santiago, J., G\'alvez, M. C. et~al.
2001, MNRAS, 328, 45

\bibitem[2007]{Mu07} Mullally, F., Kilic, M., Reach, W. T. et~al. 2007, ApJS,
171, 206

\bibitem[2004]{No04} Nordstr\"om, B., Mayor, M., Andersen, J. et~al. 2004, A\&A,
418, 989

\bibitem[1924]{Oe24} \"Opik, E. 1924, Tartu Obs. Publ. 25, No. 6

\bibitem[2002]{Or02} Ortega, V. G., de la Reza, R., Jilinski, E., Bazzanella, B.
2004, ApJ, 575, L75

\bibitem[2004]{Or04} Ortega, V. G., de la Reza, R., Jilinski, E., Bazzanella, B.
2004, ApJ, 609, 243

\bibitem[1997]{Ot97} Ottmann, R., Fleming, T. A., Pasquini, L. 1997, A\&A, 322,
785

\bibitem[2000]{Pa00} Palasi, J. 2000, Birth and Evolution of Binary Stars,
Poster Proceedings of IAU Symposium No. 200 on The Formation of Binary Stars,
held 10-15 April, 2000, in Potsdam, Germany. Edited by Bo Reipurth and Hans
Zinnecker, 2000, p. 145. 

\bibitem[1990]{Pa90} Pallavicini, R., Tagliaferri, G., Stella, L. 1990, A\&A,
228, 403

\bibitem[1997]{Pe97} Perryman, M. A. C., Lindegren, L., Kovalevsky, J. et~al.
1997, A\&A, 323, L49

\bibitem[2004]{PA04} Poveda, A., Allen, C. 2004, RMxAC, 21, 49

\bibitem[2009]{Ra09} Radigan, J., Lafreni\`ere, D., Jayawardhana, R., Doyon, R.
2009, ApJ, 698, 405

\bibitem[1998]{Re98} Rebolo, R., Zapatero Osorio, M. R., Madruga, S. et~al.
1998, Science, 282, 1309

\bibitem[2004]{Re04} Reid, I. N., Cruz, K. L., Allen, P. et~al. 2004, AJ, 128,
463

\bibitem[1982]{RK82} Retterer, J. M., King, I. R. 1982, ApJ, 254, 214

\bibitem[2007]{Ri07} Richichi, A., Fors, O., Merino, M. et~al. 2006, A\&A, 445,
1081

\bibitem[2006]{Ro06} Robinson, S. E., Strader, J., Ammons, S. M. et~al. 2006,
ApJ, 637, 1102 

\bibitem[2008]{Roe08} R\"oser, S., Schilbach, E., Schwan, H. et~al. 2008, A\&A,
488, 401

\bibitem[1992]{Ry92} Ryan, S.~G. 1992, AJ, 104, 1144

\bibitem[1989]{SG89} Saarinen, S., Gilmore, G. 1989, MNRAS, 237, 311

\bibitem[1998]{SLY98} Saffer, R. A., Livio, M., Yungelson, L. R. 1998, ApJ, 502,
394

\bibitem[2003]{SG03} Salim, S., Gould, A. 2003, ApJ, 582, 1011

\bibitem[2003]{Sc03} Scholz, R.-D., McCaughrean, M. J., Lodieu, N., Kuhlbrodt,
B. 2003, A\&A, 398, L2

\bibitem[2005]{Se05a} Seifahrt, A., Guenther, E., Neuh\"auser, R. 2005, A\&A,
440, 967 

\bibitem[1979]{Sh79} Shipman, H. L. 1979, ApJ, 228, 240

\bibitem[2005]{Si05} Siegler, N., Close, L. M., Cruz, K. L. et~al. 2005, ApJ,
621, 1023 

\bibitem[2000]{SDF00} Siess, L., Dufour, E., Forestini, M. 2000, A\&A, 358, 593

\bibitem[2006]{Sk06} Skrutskie, M. F., Cutri, R. M., Stiening, R. et~al. 2006,
AJ, 131, 1163

\bibitem[2000]{St00} Strassmeier, K., Washuettl, A., Granzer, T. et~al. 2000,
A\&AS, 142, 275 

\bibitem[1993]{SM93} Soderblom, D. R., Mayor, M. 1993, AJ, 105, 226

\bibitem[1964]{To64} Tolbert, C. R. 1964, ApJ, 139, 1105

\bibitem[2006]{To06} Torres, C. A. O., Quast, G. R., da~Silva, L. et~al. 2006,
A\&A, 460, 695

\bibitem[1995]{vA95} van~Altena, W. F., Lee, J. T., Hoffleit, E. D. 1995, 
The general catalogue of trigonometric [stellar] parallaxes, 
New Haven, CT: Yale University Observatory, 1995, 4th~ed.

\bibitem[2007]{vL07} van Leeuwen, F. 2007, A\&A, 474, 653

\bibitem[2006]{VV06} Violat-Bordonau, F., Violat-Mart\'{\i}n, D. 2006, Open
European Journal on Variable Stars, 53, 1

\bibitem[1917]{Vo17} Vo\^ute, J. 1917, MNRAS, 77, 650

\bibitem[1991]{WS91} Wasserman, I., Weinberg, M. D. 1991, ApJ, 382, 149

\bibitem[1988]{WS88} Weinberg, M. D., Wasserman, I. 1988, ApJ, 329, 253

\bibitem[1987]{We87} Weinberg, M. D., Shapiro, S. L., Wasserman, I. 1987, ApJ,
312, 367

\bibitem[1984]{We84} Weis, E.~W. 1984, ApJS, 55, 289

\bibitem[1988]{We88} Weis, E.~W. 1984, AJ, 96, 1710

\bibitem[2008]{We08} West, A. A., Hawley, S. L., Bochanski, J. J. et~al. 2008,
AJ, 135, 785

\bibitem[2006]{WL06} Wertheimer, J. G., Laughlin, G. 2006, AJ, 132, 1995

\bibitem[2004]{ZOM04} Zapatero Osorio, M. R., Mart\'{\i}n, E. L. 2004, A\&A,
419, 167

\bibitem[2004]{ZS04} Zuckerman, B., Song, I. 2004, ARA\&A, 42, 685

\bibitem[2001]{Zu01} Zuckerman, B., Song, I., Bessell, M. S., Webb, R. A. 2001,
ApJ, 562, L87

\bibitem[2003]{Zu03} Zuckerman, B., Koester, D., Reid, I. N., H\"unsch, M. 2003,
ApJ, 596, 477

\end{thebibliography}
\end{document}